\documentclass{aa}  

\usepackage{graphicx}
\usepackage{txfonts}
\usepackage[amssymb,mediumspace,thickqspace]{SIunits}
\usepackage{bm}
\usepackage{subcaption}

\begin{document}

   \title{Spread of the dust temperature distribution in circumstellar
   disks}


   \author{S. Heese
          \inst{1}\and
          S. Wolf\inst{1}\and A. Dutrey\inst{2}\and S. Guilloteau\inst{2}
          }
   \institute{Institute for Theoretical Physics and Astrophysics, University of 
Kiel,
               Leibnizstraße 15, 24118 Kiel, Germany
             \and Université Bordeaux 1, Laboratoire d’Astrophysique de 
Bordeaux 
              (LAB), UMR 5804, 2 rue de l’Observatoire, BP 89,
              33270 Floirac Cedex, France
             }
             
\abstract{Accurate temperature calculations for circumstellar disks are 
particularly important for their chemical evolution. 
Their temperature distribution is determined by the optical properties of the 
dust grains, which,
among other parameters, depend on 
their radius. However, in most disk studies, only average optical properties and 
thus
an average temperature is assumed to account for an ensemble
of grains with different radii.}
{We investigate the impact of subdividing the grain radius distribution into 
multiple sub-intervals
on the resulting dust temperature distribution and spectral energy 
distribution (SED).}
{The temperature distribution, the relative grain surface below a certain 
temperature, the freeze-out radius, and 
the SED were computed 
for two different scenarios:
(1) Radius distribution represented by 16 logarithmically distributed radius 
intervals, and
(2) radius distribution represented by a single grain species with averaged 
optical properties (reference).}
{Within the considered parameter range, i.e., of grain radii between 
$5\,\mathrm{\nano \metre}$ and $1\,\mathrm{\milli \metre}$ and an optically 
thin and thick disk with a parameterized density distribution, we obtain the 
following results:
In optically thin disk regions, the temperature spread can be as large as 
$\sim63\%$ and the relative grain surface
below a certain temperature is
lower than in the reference disk. With
increasing optical depth, the difference in the midplane temperature and the 
relative grain surface below a certain temperature
decreases. Furthermore, below $\sim 20\,\mathrm{\kelvin}$, this fraction is 
higher for the reference disk than for the 
case of multiple grain radii,
while it shows the opposite behavior for temperatures above this threshold. 

The thermal emission in the case of
multiple grain radii at short wavelengths is
stronger than for the reference disk.
The freeze-out radius (snowline) is a function of grain radius, spanning a 
radial range between the coldest and warmest grain species of $\sim 
30\,\mathrm{AU}$.}
{}

\maketitle

\section{Introduction}

The dust phase of the interstellar matter (ISM) shows a broad range of grains 
with different radii.
In circumstellar disks around young stellar objects, which form during the 
collapse of the densest parts of the ISM, 
this dust phase is modified by various processes 
(\citealt{2013A&A...553A..69G}); this results in a potentially
even larger spread of grain radii, which is due to the process of dust grain
growth.
Even though the dust amounts to only $1\%$ of the disk mass, it dominates the 
extinction of the disk. Consequently, understanding its properties including the 
impact of the 
existence of grains with various radii is fundamental for understanding 
circumstellar disk physics.

Constraints on the properties of the dust phase can be derived from 
multiwavelength observations
of circumstellar disks and subsequent radiative transfer modeling
(e.g., \citealt{2012A&A...543A..81M,2013A&A...553A..69G}).
Because of computational constraints, usually only a single dust phase is 
assumed that represents 
average optical properties, resulting in an average dust temperature.
Whether this simplification is justified was first investigated by 
\cite{2003ApJ...582..859W}.
It was found that the temperature dispersion at the inner boundary of a dust 
shell can amount to $\gg 100\,\mathrm{\kelvin}$
and has to be considered in, for example, chemical networks. However, 
in the 
same study the difference 
between SEDs based on a real dust grain
distribution and mean dust grain parameters was 
found to be below 10\,\% if $2^5-2^6$ radius bins are considered.
Since this early study the level of constraints provided by disk observations 
has increased significantly over the past decade.
 As the accuracy of the observations increases, the 
demand in accuracy of the numerical models increases as well. Also, because the 
computing power has increased,
a more correct treatment of the grain radius distribution becomes 
feasible.

The aim
of our study is to assess the effects of a correct treatment of the grain 
radius distribution.
This study is motivated by the fact that an
accurate calculation of dust and gas temperatures is of particular importance 
for the astrochemical evolution of 
circumstellar disks.
For example, below $20\,\mathrm{\kelvin}$,
CO freezes out, water below $100\,\mathrm{\kelvin}$ 
(\citealt{2010ApJ...716..825O}).
\cite{2008A&A...488..565C} suggested that in the CQ Tau disk, where an 
important 
depletion of CO is observed despite a warm temperature, CO may remain trapped 
onto large grains that are cold enough to prevent thermal CO desorption.   
When CO is frozen out, deuterium fractionation is very efficient 
(\citealt{2015arXiv151202986P})
and the depletion of CO and similar molecules affect the cooling and therefore 
the evolution 
of star-forming clouds (\citealt{2016MNRAS.456.2586H}).
Likewise, the dominant carbon bearing entity depends on the temperature 
structure
and age of the disk. According to \cite{2015A&A...579A..82R}, CO trapped
on grain surfaces is the most abundant C-bearing species between
$\sim5\,\mathrm{\kelvin}$ and $\sim 15\,\mathrm{\kelvin}$.
 
Furthermore, \cite{2016A&A...586L...1G} proposed the 
difference in temperature between larger and smaller grains as a possible 
explanation for the 
low dust temperature of the Flying saucer, derived from measurements of the 
velocity gradients due to
Keplerian rotation and variations in the CO background as a function of 
velocity.

Using a dust distribution consisting of astronomical silicate 
and graphite,
the difference between a more precise treatment of the grain radius 
distribution, 
considering individual
grains of different radii, taking their relative abundances into account,
and using average optical properties for the whole radius range is assessed for 
the following quantities: 
(1) The spatial temperature distribution, (2) the relative dust grain
surface below a certain temperature, (3) the freeze-out radius and (4) the 
spectral energy distribution (SED). 

In $\S2$, we briefly introduce the applied radiative transfer (RT) code, the 
disk model, and dust grain model.
In $\S3.1$, we discuss the results for an optically thin disk. Subsequently, we 
extend the discussions to the 
peculiarities of disks with an optically thick midplane
 ($\S3.2$). In $\S$3.3, we discuss the influence of more 
realistic
disk masses, an inner cavity ($\S$3.4), and having larger disks 
($\S$3.5).
Finally, we apply our results to the special case of the 
Flying Saucer in $\S$3.6.

\section{Setup}

In this section, we briefly describe the applied Monte Carlo (MC) radiative 
transfer
code and the disk model.

\subsection{Monte Carlo radiative transfer code and disk setup}
\label{seq:setup}

The simulations presented in this article
are performed with the 3D
dust radiative transfer code Mol3D (\citealt{Ober2015}).
This program calculates the dust temperature based on the 
assumption of local thermal equilibrium. The calculation of the spatial dust 
temperature 
distribution combines the continuous absorption method proposed by 
\cite{1999A&A...344..282L}
and immediate temperature correction by \cite{2001ApJ...554..615B}.

We apply a parameterized density distribution according to 
\cite{1973AA....24..337S},
successfully used in earlier studies of young circumstellar disks (e.g.,
\citealt{2003ApJ...588..373W, 2012A&A...543A..81M, 2013A&A...553A..69G}) as 
follows:
\begin{align}
 &\rho(r, z)\sim \left(\frac{r}{100\,\mathrm{AU}}\right)^{-\alpha}\cdot 
\exp\left(-\frac{1}{2}\frac{z^2}{h(r)^2}\right) \label{equ:rho} \text{, with}\\
&h(r)=h_{100}\cdot \left(\frac{r}{100\,\mathrm{AU}}\right)^{\beta}, 
\label{equ:h}
\end{align}
$\alpha=2.625$, $\beta=1.125$ and 
$h_{100}=10.0\,\mathrm{AU}$. The condition of hydrostatic equilibrium 
is not considered in this approach. For the inner and outer radius, we use 
$1\,\mathrm{AU}$
and $120\,\mathrm{AU}$, respectively. 
We set the inner radius ($1\,\mathrm{AU}$) outside the dust sublimation radius 
mainly because
the dust sublimation radius depends on the dust grain radius. Thus, calculate
the dust sublimation radius self-consistently would require an iterative 
approach.
Furthermore, a grain-radius dependent sublimation radius would unnecessarily 
complicate the comparison with the reference disk.

For the dust mass, we assess two extreme
cases: An optically thin disk, which is optically thin even for the shortest 
wavelength, and a disk
with an optically thick midplane, 
where the midplane is optically thick even for the longest wavelength of our 
calculations.
In the optically thin case, the mass is irrelevant as long as the optical depth 
for the shortest wavelength
is much smaller than 0.1. For the optically thick case, we use a dust mass of 
$M_\text{Dust}=10^{-3}\,\mathrm{M_{\odot}}$.
Together with the gas, the total mass of the disk amounts to 
$10^{-1}\,\mathrm{M_{\odot}}$.
As the central radiation
source we use a T Tauri star with $M_{\star}=0.7\,\mathrm{M_{\odot}}$, 
$R_{\star}=2.0\,\mathrm{R_{\odot}}$ and $T_\text{eff}=4000\,\mathrm{\kelvin}$.
To calculate the observational quantities, we assume a distance of 
$140\,\mathrm{pc}$
and a disk inclination of $0^\circ$.

 \subsection{Dust}
 \label{seq:dust}
 
The dust in our model consists of a homogeneous mixture of graphite and 
astronomical silicate
adapted from \cite{2001ApJ...548..296W}. The dust mixture has a density of
$\rho=2.5\,\mathrm{g/cm^3}$ and consists of $62.5\,\mathrm{\%}$ silicate and
$37.5\,\mathrm{\%}$ graphite. Because of the anisotropic optical properties of 
graphite, we use 
the $\frac{1}{3}$ - $\frac{2}{3}$ approximation by \cite{1993ApJ...414..632D}, 
where the extinction coefficient $Q_\text{ext}$
is calculated as follows:
\begin{align}
Q_\text{ext}=\frac{1}{3}Q_\text{ext}(\epsilon_{\parallel})+\frac{2}{3}Q_\text{
ext}(\epsilon_{\perp}).
\end{align}

We apply a grain radius distribution following $n(a)\sim a^{-3.5}$ with a 
lower and upper grain radius
of $5\,\mathrm{\nano \metre}$ and $1\,\mathrm{\milli \metre}$, respectively.
The lower limit is motivated by the constraints derived for the dust in the ISM
(\citealt{1977ApJ...217..425M}), which is also consistent with the analysis of 
dust in protoplanetary disks. 
The upper limit is consistent with the evidence of grains of 
millimeter-to-centimeter radii in the inner dense regions of these disks (e.g., 
\citealt{2013A&A...553A..69G}).
This range of radii is subdivided into 16
logarithmically distributed intervals. In each of these intervals weighted mean 
dust grain parameters are computed
(for the procedure see below).
A list of the intervals is shown in Tab.\,\ref{tab:sizes}. This dust setup is 
referred to as MGS (multiple grain size) in the following.

As reference, we compute the weighted mean dust grain parameters taking 
the optical properties of 
the entire range of grain radii into account.
The weighted mean values are
derived as follows
(\citealt{1978cdii.book.....M, 1980AcMPh..21...19S}):
\begin{align}
 \langle C_\text{ext}\rangle =& 
\sum_{j=1}^{n_\text{D}}\int_{a_\text{min}}^{a_\text{max}}w\!_j(a)C_{\text{ext}_j
}(a)da, \\
 \langle C_\text{abs}\rangle =& 
\sum_{j=1}^{n_\text{D}}\int_{a_\text{min}}^{a_\text{max}}w\!_j(a)C_{\text{abs}_j
}(a)da, \\
 \langle C_\text{sca}\rangle =& 
\sum_{j=1}^{n_\text{D}}\int_{a_\text{min}}^{a_\text{max}}w\!_j(a)C_{\text{sca}_j
}(a)da, \\
 \langle \hat{S}\rangle =& 
\sum_{j=1}^{n_\text{D}}\int_{a_\text{min}}^{a_\text{max}}w\!_j(a)\hat{S}
\!_j(a)da,
\end{align}
where $w\!_j(a)$ is the weight of the $j$th component of the dust 
grain mixture, 
$n_\text{D}$ is the number of dust species, and $C_\text{abs}$ 
and 
$C_\text{ext}$ are the absorption and extinction cross sections.
Furthermore, $\hat{S}\!_j$ is the Mueller matrix, which is used to 
describe
the modification of the Stokes vector of the photon due to the interaction of a
photon with the absorbing and scattering medium.
The weight of each component results from its abundance with respect to its dust 
grain number density and the 
radius distribution of the respective material (\citealt{2003ApJ...582..859W}),
\begin{align}
 \sum_{j=1}^{n_\text{D}}\int_{a_\text{min}}^{a_\text{max}}w\!_j(a)da=1.
\end{align}

 \subsection{Absorption cross section}
 \label{seq:qabs}
 
 The temperature distribution of the grains depends on their absorption
 cross sections.
 The star emits its maximum radiation at
 $\sim 750\,\mathrm{\nano \metre}$, while the grains radiate between
 $\sim7\,\mathrm{\micro \metre}$ and $\sim120\,\mathrm{\micro \metre}$, 
 derived using Wien's displacement law
 from the calculated temperatures ($\sim25\,\mathrm{\kelvin}$ to $\sim 
450\,\mathrm{\kelvin}$).
 Thus, a large absorption cross section around $750\,\mathrm{\nano \metre}$ 
increases
 the efficiency of heating of the grains, while a large absorption cross section
 between $7\,\mathrm{\micro \metre}$ and $120\,\mathrm{\micro \metre}$ 
 increases the efficiency of cooling.
 In the following, the expected temperature differences for grains in different 
radius bins
 are discussed in more detail:
 
 \begin{figure}
  \centering
  \begin{subfigure}[b]{1.0\columnwidth}
   \includegraphics[width=1.0\columnwidth]{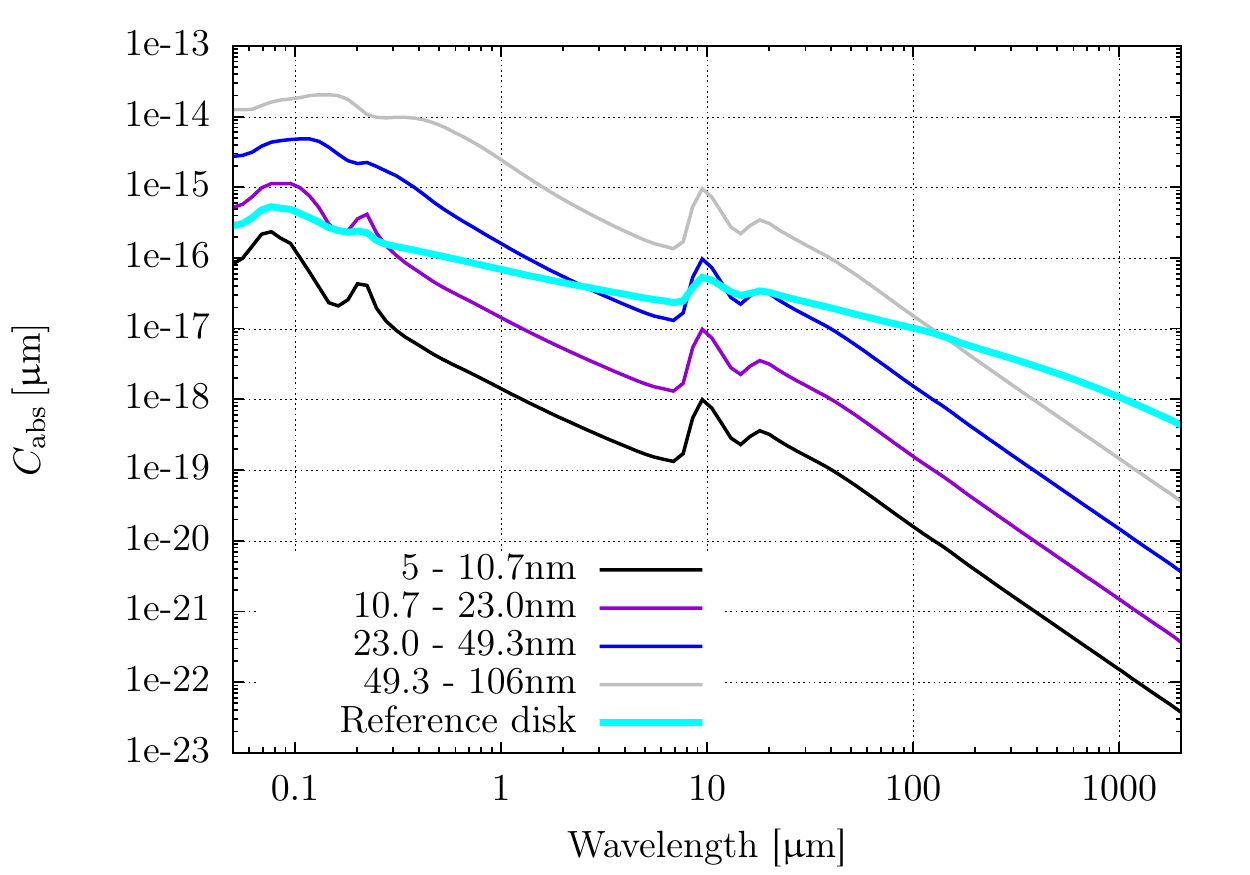}
  \end{subfigure}
\begin{subfigure}[b]{1.0\columnwidth}
 \includegraphics[width=1.0\columnwidth]{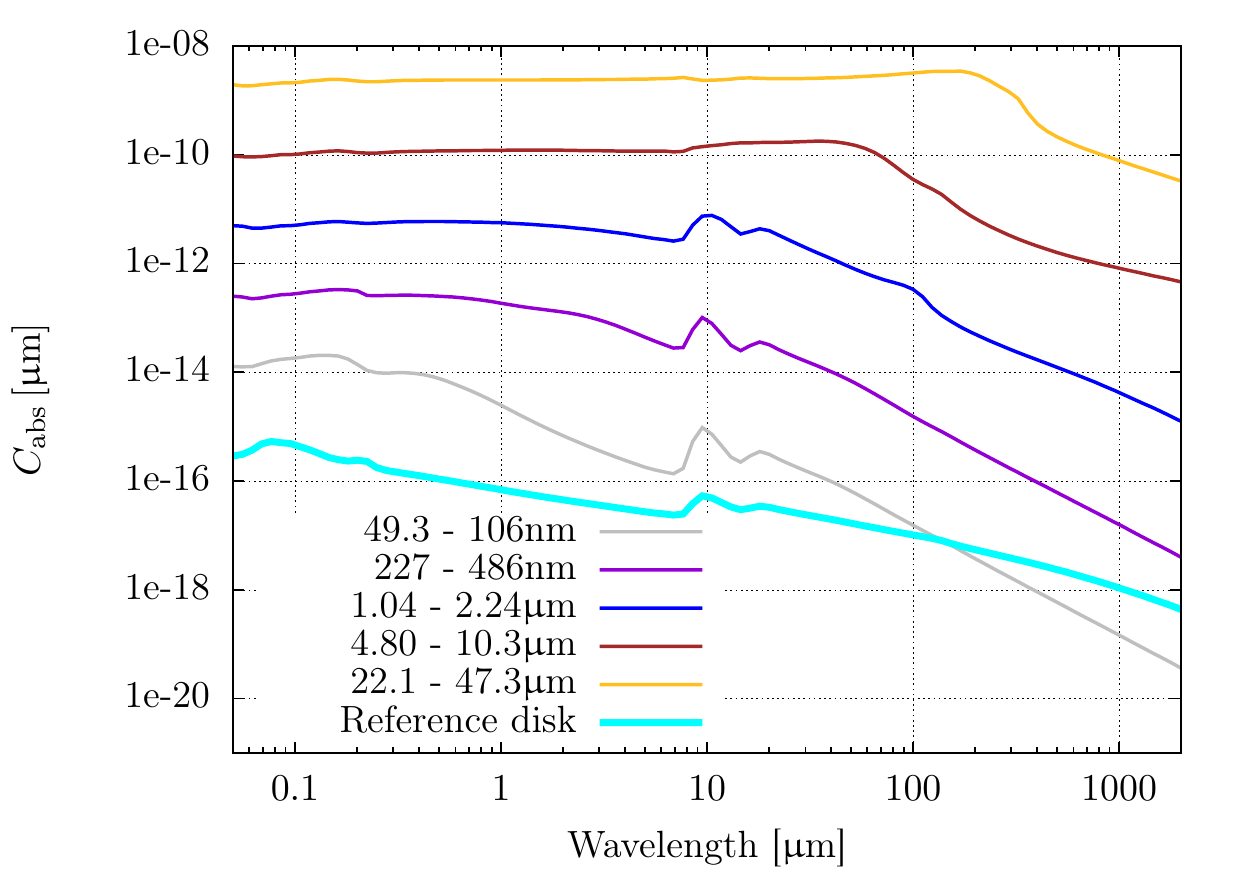}
\end{subfigure}
\begin{subfigure}[b]{1.0\columnwidth}
  \includegraphics[width=1.0\columnwidth]{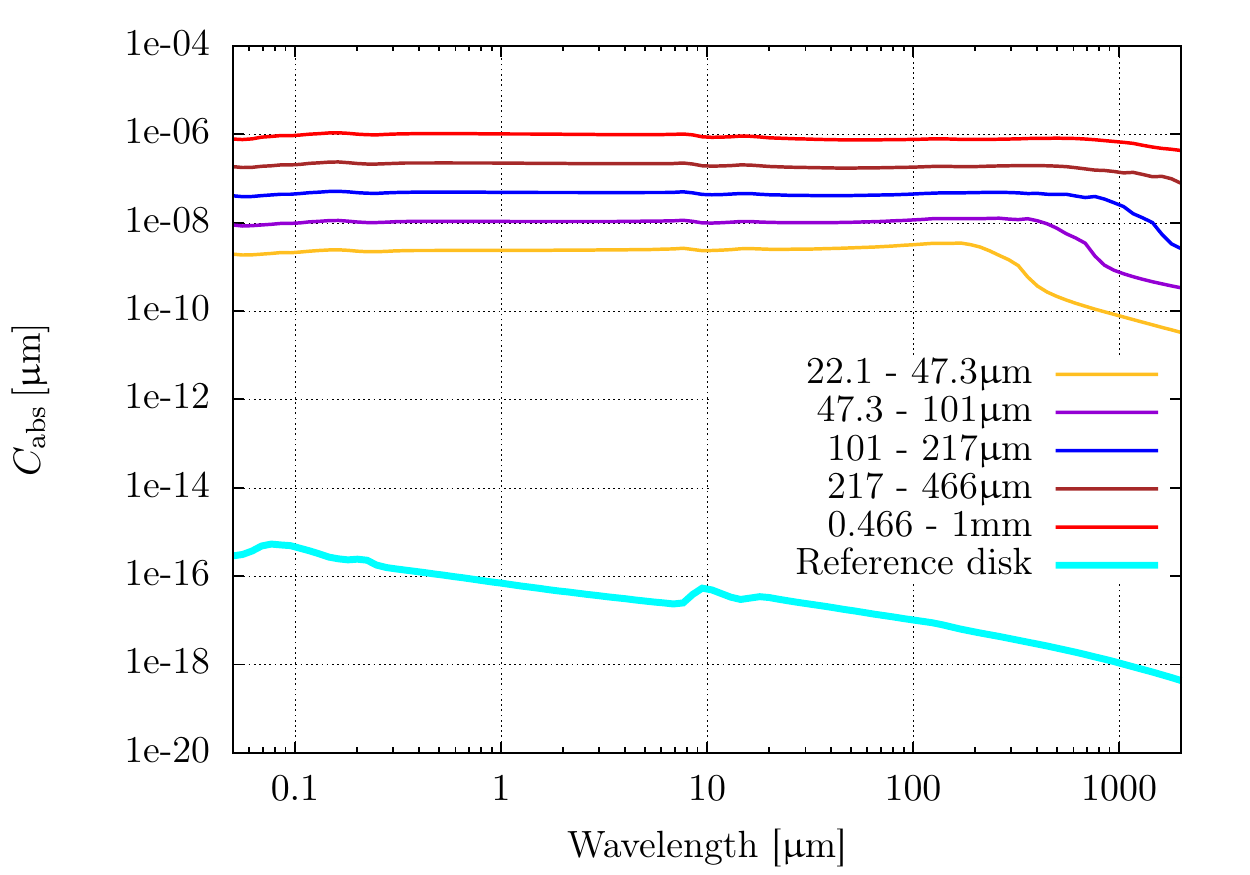}
  \label{img:Qabs1e20_3}
\end{subfigure}
\caption{Absorption cross sections for the different dust grain species. For 
details, see Sect.\,\ref{seq:qabs}.}
\label{img:Cabs}
 \end{figure}

  \begin{itemize}
 \item The absorption cross sections for small grains ($5\,\mathrm{\nano 
\metre}$ to $106\,\mathrm{\nano \metre}$)
 increase faster for shorter than for longer wavelengths (see 
Fig.\,\ref{img:Cabs}, top). 
 Consequently, the grain temperature should increase with increasing grain
 radius.
  \item The absorption cross sections for intermediate sized grains 
($50\,\mathrm{\nano \metre}$ to $47\,\mathrm{\micro \metre}$)
 are shown in Fig.\,\ref{img:Cabs}, middle.
The difference in the cross sections for short and long wavelengths decreases.
 Consequently, the grain temperature should decrease with increasing grain 
radius. However, for the grain species with radii between 
$4.8\,\mathrm{\micro\metre}$ and $10\,\mathrm{\micro\metre}$, the absorption 
cross section has a maximum between $\sim 7\,\mathrm{\micro\metre}$ and 
$70\,\mathrm{\micro\metre}$, which does not exist for larger 
grains. Thus, for grain temperatures between $41\,\mathrm{\kelvin}$ and 
$410\,\mathrm{\kelvin}$, the grain species with radii between 
$4.8\,\mathrm{\micro\metre}$ and $10\,\mathrm{\micro\metre}$ has the 
most 
efficient cooling.

 \item The absorption cross sections for the largest grains 
($22\,\mathrm{\micro \metre}$ to $1\,\mathrm{\milli \metre}$)
 slightly increase with wavelengths for grains with radii between 
$22\,\mathrm{\micro\metre}$ and $47\,\mathrm{\micro\metre}$ (see 
Fig.\,\ref{img:Cabs}, bottom). For larger grains, this increase 
becomes weaker 
and 
changes into a decrease for grains larger than $217\,\mathrm{\micro\metre}$.
 Thus, the grain temperature should increase again with increasing grain radius.
 \end{itemize}
 For comparison, the absorption cross section of the reference grains is shown 
in the same figures.
 
 \section{Results}

Because of the wide range of optical depths and different heating processes 
involved, such as direct stellar radiation, scattered light, and thermal 
re-emitted radiation, the
temperature structure in protoplanetary disks shows a complex structure. 
In $\S3.1$, we present results for an optically thin disk, which are applicable 
for the upper layers of protoplanetary disks, while the results for a disk
with an optically thick midplane
($\S3.2$) are applicable in the midplane of these disks.
In
$\S3.3$ to 
$\S3.5$ we extend our study to more realistic disks with different masses, 
inner cavities of different sizes and 
various outer radii. 
In $\S3.6$, we apply the MGS grain radius distribution to the special 
case of 
the Flying Saucer.

 \subsection{Optically thin case}
 
 In this case, all dust grains are directly heated by the central star.
 The contribution by scattered and re-emitted radiation on the spatial 
 dust temperature distribution is negligible.
 
  \subsubsection{Temperature distribution}
  \label{seq:mtd1}

\begin{figure}
  \centering
  \includegraphics[width=1.0\columnwidth]{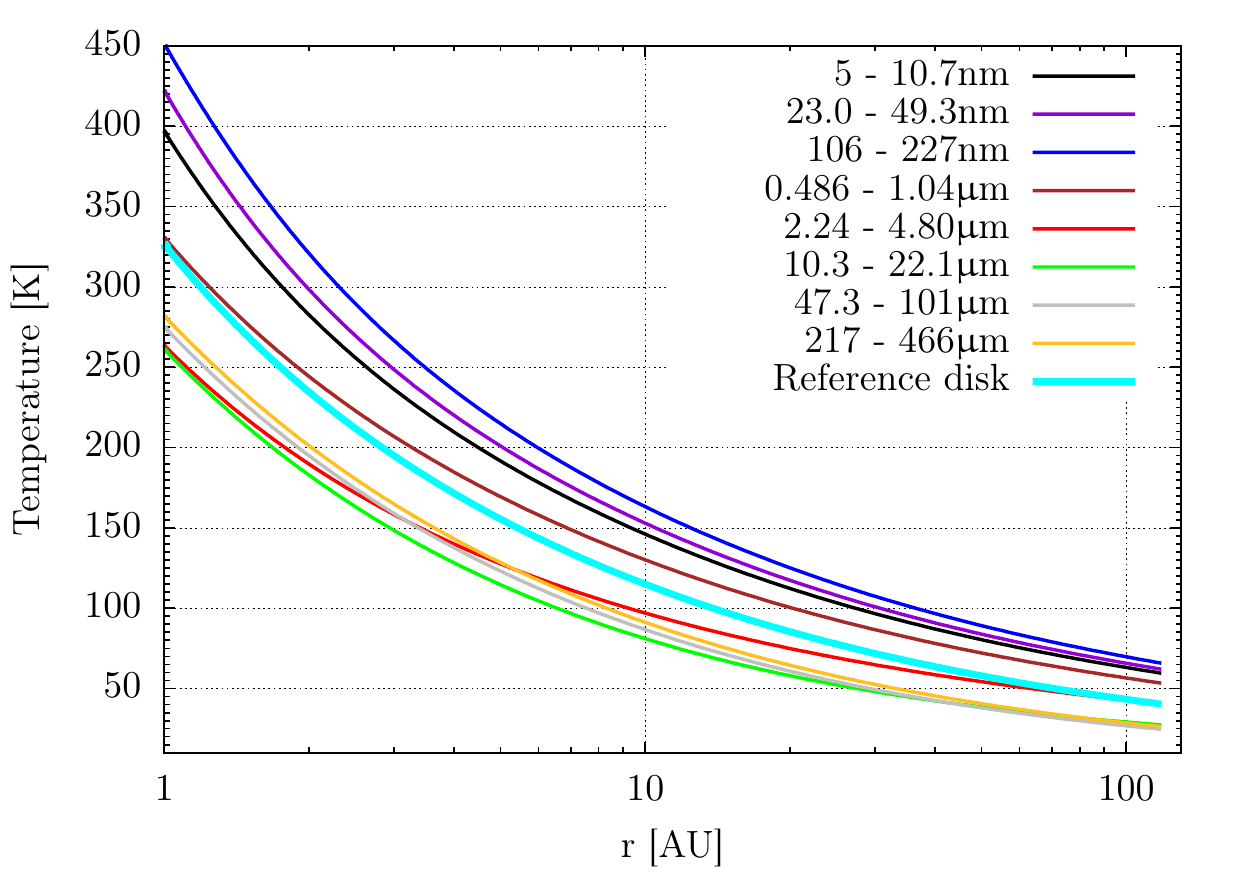}
    \caption{Radial temperature distribution in the optically thin case. For 
details, see Sect.\,\ref{seq:mtd1}.}
  \label{img:temp1e20}
 \end{figure}

\begin{figure}
  \centering
  \begin{subfigure}[b]{1.0\columnwidth}
   \includegraphics[width=1.0\columnwidth]{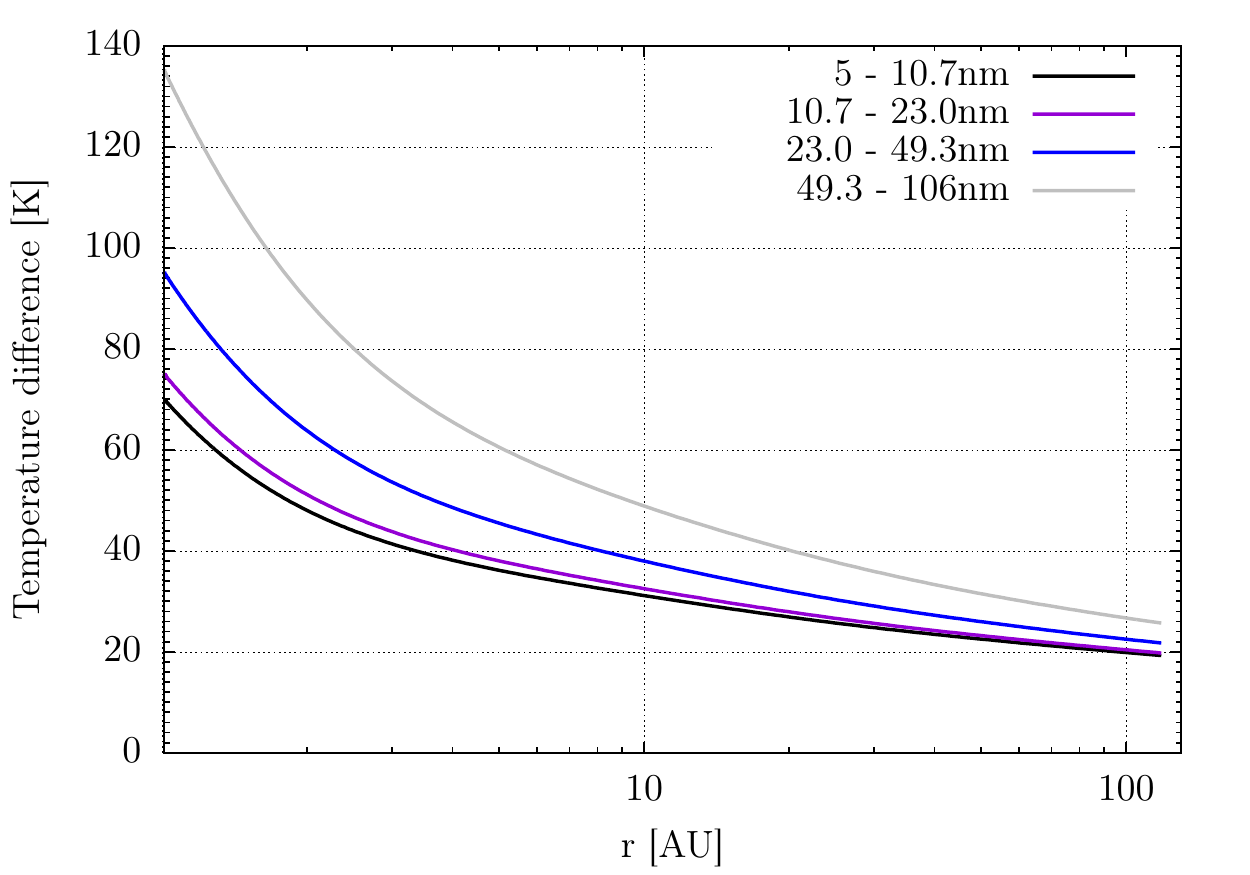}
  \end{subfigure}
\begin{subfigure}[b]{1.0\columnwidth}
 \includegraphics[width=1.0\columnwidth]{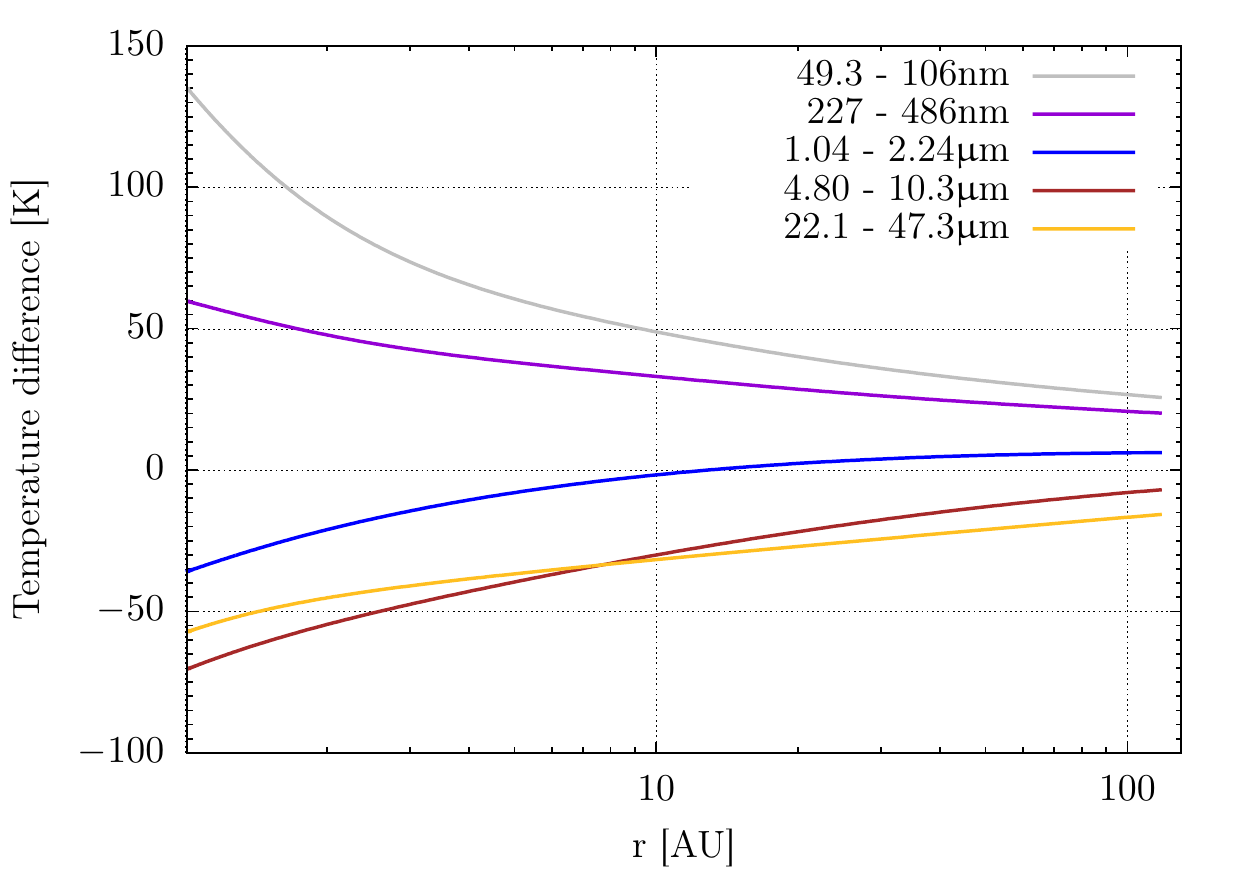}
\end{subfigure}
\begin{subfigure}[b]{1.0\columnwidth}
  \includegraphics[width=1.0\columnwidth]{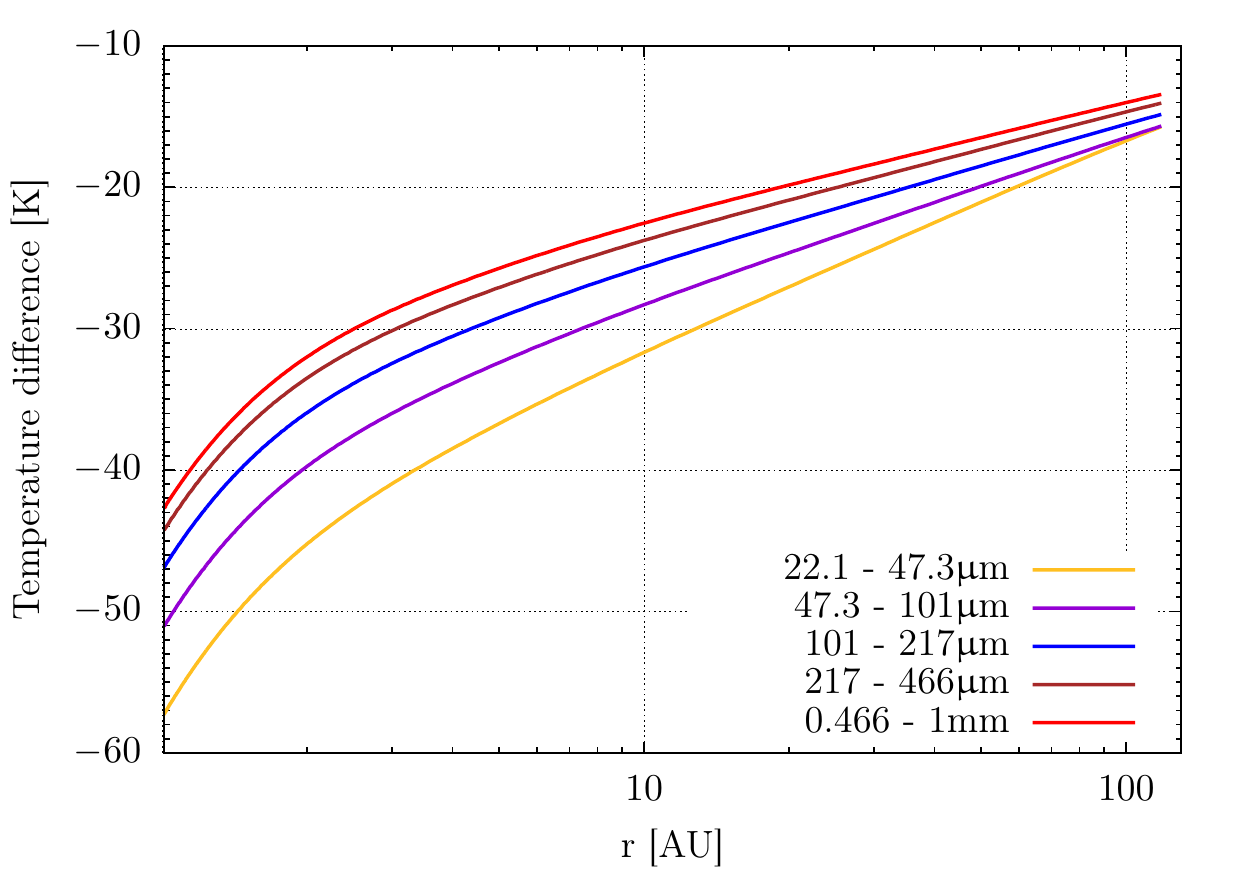}
\end{subfigure}
\caption{Temperature differences
from the reference disk in the optically thin case. For details, see 
Sect.\,\ref{seq:mtd1}.}
\label{img:diff1e20}
 \end{figure}

 Here, we investigate the temperature
 distribution to see if the predictions based on the absorption cross sections 
hold (see Fig.\,\ref{img:temp1e20}).
 To study the details of the spread, the differences from the reference
 disk are shown in Fig.\,\ref{img:diff1e20}.
 The differences between grains of different radii reach values of up 
to 
$\sim200\,\mathrm{\kelvin}$, which correspond to $45\,\%$ of the 
 temperature value of the hottest grains. For lower temperatures further away 
from the star, the differences can even reach $63\,\%$ of the 
temperature 
 value of the hottest grains.

  The 
 temperature first increases with increasing grain radii up to a radius of 
 $\sim100\,\mathrm{\nano \metre}$, then decreases with further increasing radii 
up to a radius
 between $10\,\mathrm{\micro \metre}$ and $47\,\mathrm{\micro \metre}$
 before the temperature increases again.
 This behavior is directly related to the wavelength 
 dependence of the absorption cross sections
 $C_\text{abs}$ shown in Fig.\,\ref{img:Cabs}.
 
 \subsubsection{Grain surface}
 \label{seq:gs1}
 
\begin{figure}
  \centering
  \includegraphics[width=1.0\columnwidth]{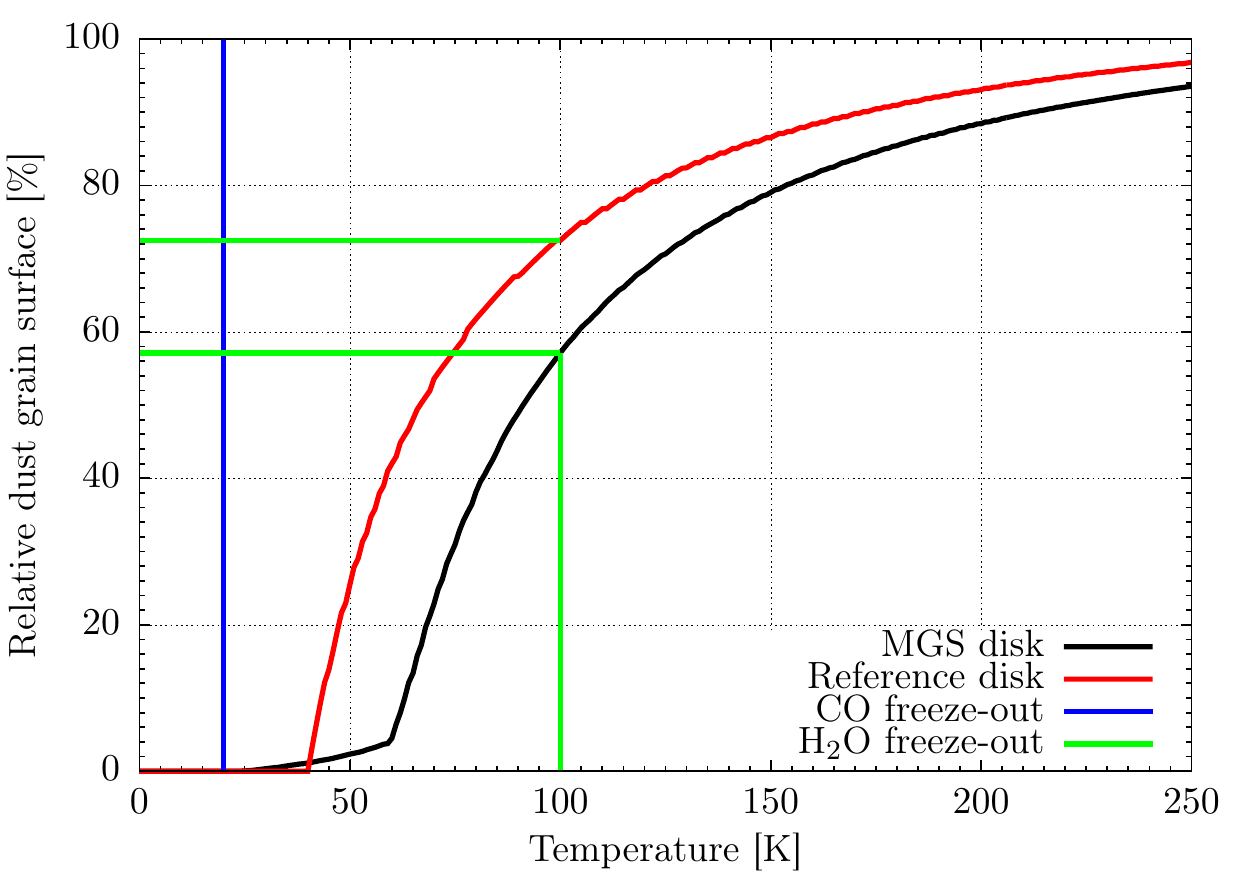}
  \caption{Relative dust grain surface below a certain temperature in the 
optically thin case. For details, see 
  Sect.\,\ref{seq:gs1}.}
  \label{img:gs1e20}
 \end{figure}

 Besides the temperature, the net surface of the dust grains is an important 
quantity that determines the rate of chemical reactions.
 For this reason, the relative dust grain surface below a certain temperature 
is now compared
 to the results for the reference disk (Fig.\,\ref{img:gs1e20}).
 We find that this surface fraction grows faster for the reference disk than it 
does for the MGS disk. However, because
 the grains with radii larger than $\sim 10\,\mathrm{\micro \metre}$ have 
temperatures between $24\,\mathrm{\kelvin}$
 and $28\,\mathrm{\kelvin}$ at the outer parts of the disk, there is dust grain 
surface available below 
 $40\,\mathrm{\kelvin}$ for the MGS disk. As the large dust grains do not 
contribute significantly to the total dust grain surface and 
 as the small dust grains have higher temperatures (over $50\,\mathrm{\kelvin}$ 
for grains with radii below
 $\sim 1\,\mathrm{\micro \metre}$), the relative dust grain surface begins to 
grow fast only above these higher temperatures.
 
 For the reference disk, the relative dust grain surface
 below $100\,\mathrm{\kelvin}$ (freeze-out temperature of water; 
\citealt{2010ApJ...716..825O}) amounts to
 $72.5\%$, while it only amounts to $57.1\%$ for the MGS disk, and the relative 
dust grain surface between $20\,\mathrm{\kelvin}$
 and $30\,\mathrm{\kelvin}$ amounts to $0.34\,\%$ for the MGS disk and 
$0\,\%$ 
for the reference disk.
 Therefore, the relative fraction of dust grain surface
 below a certain temperature (and thus available for certain reactions) is 
significantly lower in the case of a correct
 treatment of the distribution of grain radii and calculation of the 
corresponding temperature distribution.

\begin{figure}
  \centering
  \begin{subfigure}[b]{1.0\columnwidth}
   \includegraphics[width=1.0\columnwidth]{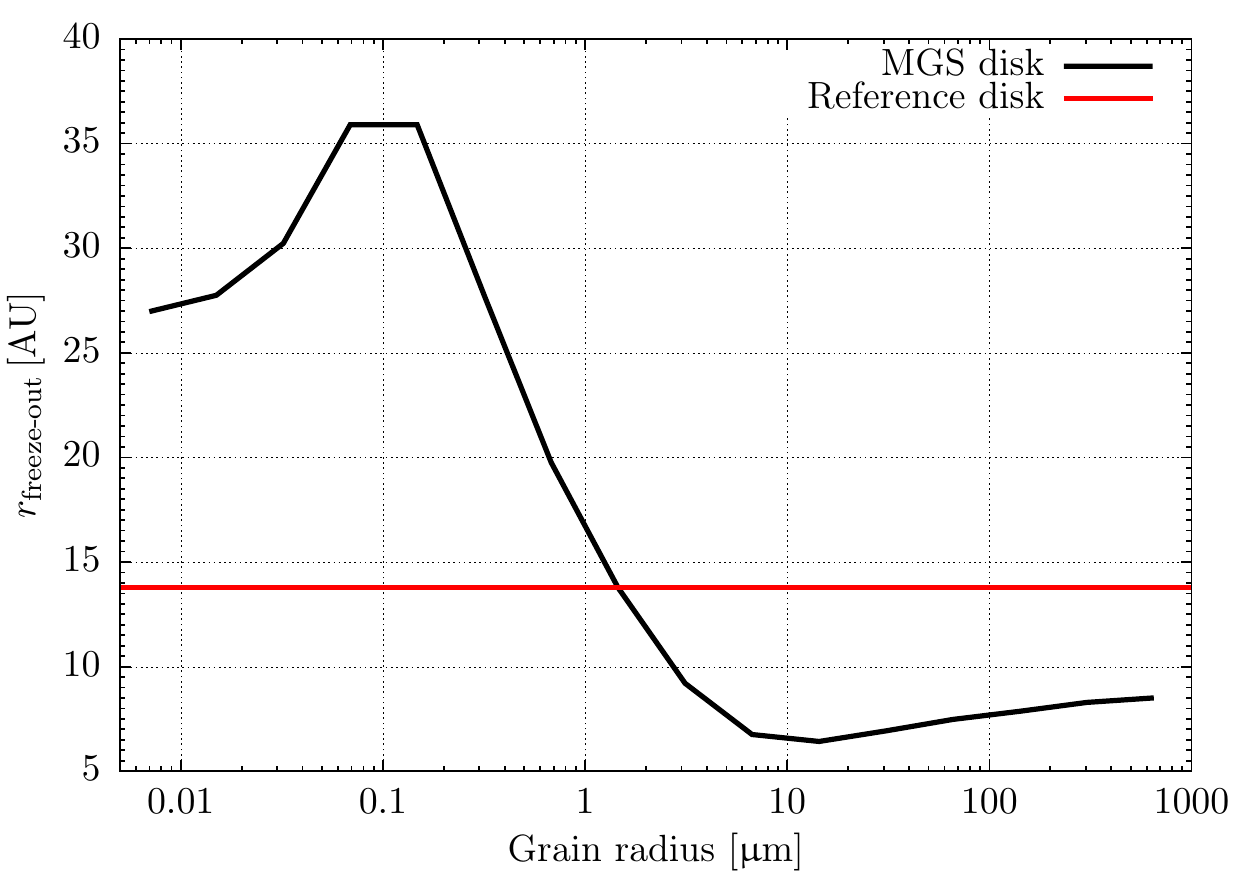}
  \end{subfigure}
\begin{subfigure}[b]{1.0\columnwidth}
\centering
   \includegraphics[width=1.0\columnwidth]{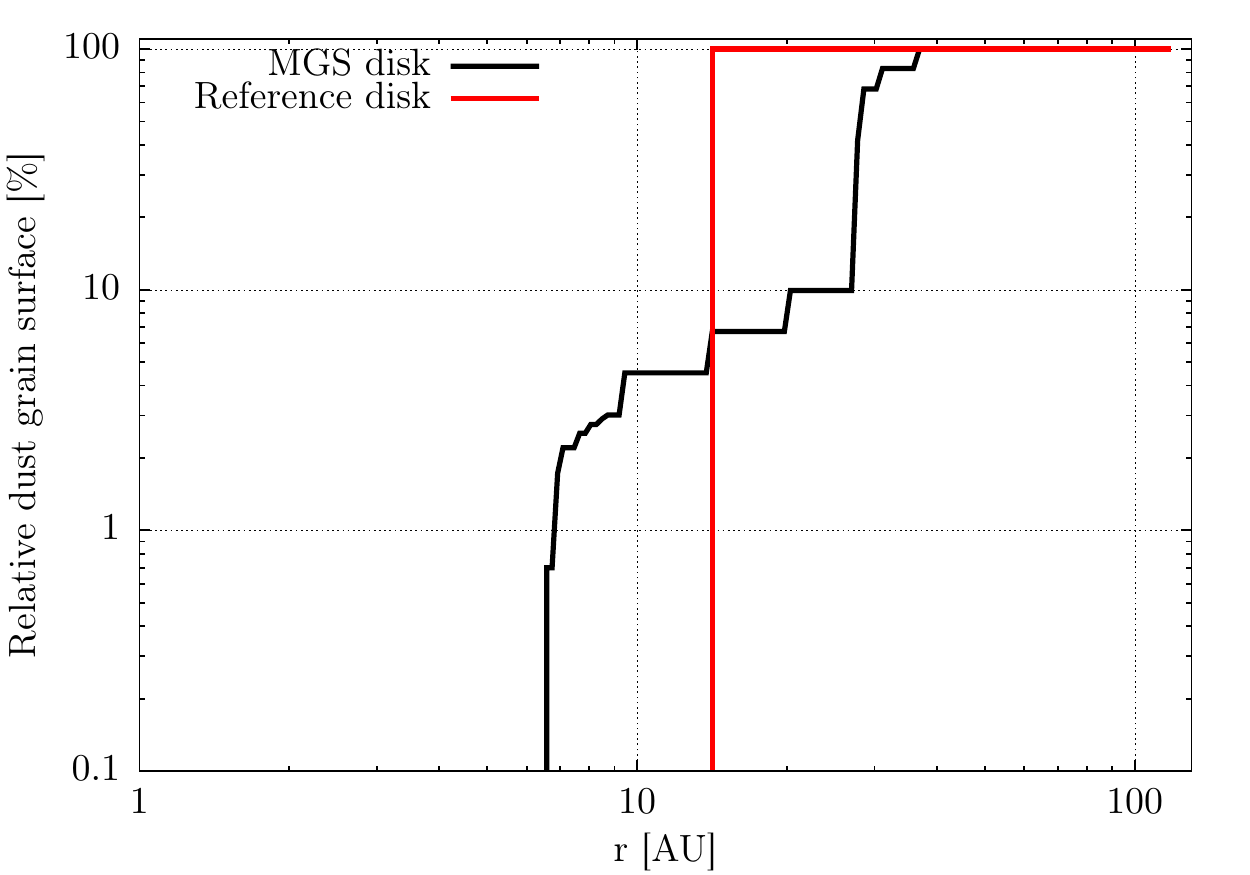}
  \end{subfigure}
  \caption{Top: Radial distance from the star, outside of which all 
grains are
  below the freeze-out temperature of water ($100\,\mathrm{\kelvin}$),
  depending on the grain radius in the 
  optically thin case. \\
 Bottom: Relative dust grain surface below the freeze-out 
temperature of water
  as a function of radial distance from the star in the optically thin case.
  For details, see Sect.\,\ref{seq:gs1}.
  \label{img:freezeout1}
}
\end{figure}

 A different perspective on this problem is to investigate the radial distance 
from the star,
 outside of which all grains of a certain radius have temperatures below the 
freeze-out temperature of a given volatile species,
 the so-called freeze-out radius $r_\text{freeze-out}$. 
In Fig.\,\ref{img:freezeout1}, top this quantity is shown for the case 
of water. 
 It first grows with increasing grain radius up to a grain radius of
  $\sim0.1\,\mathrm{\micro\metre}$, then decreases but increases again for 
grains above $\sim 20\,\mathrm{\micro\metre}$.
  This behavior is directly related to the temperatures of the different dust 
grains (see Fig.\,\ref{img:temp1e20})
  where the temperature first grows up to a grain radius of $106\,\mathrm{\nano 
\metre}$, but decreases
  to a grain radius between $4.8\,\mathrm{\micro \metre}$ and 
$47\,\mathrm{\micro \metre}$, before 
  the temperature increases again.
 
An associated quantity is the relative dust grain surface with a temperature 
below or equal to the freeze-out temperature
for all grains at a given radial distance from the star 
(see Fig.\,\ref{img:freezeout1}, bottom). Here again, only the 
fraction below the 
freeze-out temperature of water is shown as there are
no grains with temperatures below the freeze-out of CO. The result indicates 
that for the reference disk, 
the dust has a temperature below $100\,\mathrm{\kelvin}$ outside
of $\sim 14\,\mathrm{AU}$, while for the MGS disk
the transition is a bit smoother because the grains with different radii have 
different temperatures.
However, owing to the low optical depth, all grains in a certain radius bin 
and at a certain distance from the star have the same
temperature. Thus, the freeze-out on all of them happens at the same distance, 
resulting in the various steps of
this graph.

 \subsubsection{Thermal emission of the disk}
 \label{seq:rt1}
 
\begin{figure}
  \centering
  \includegraphics[width=1.0\columnwidth]{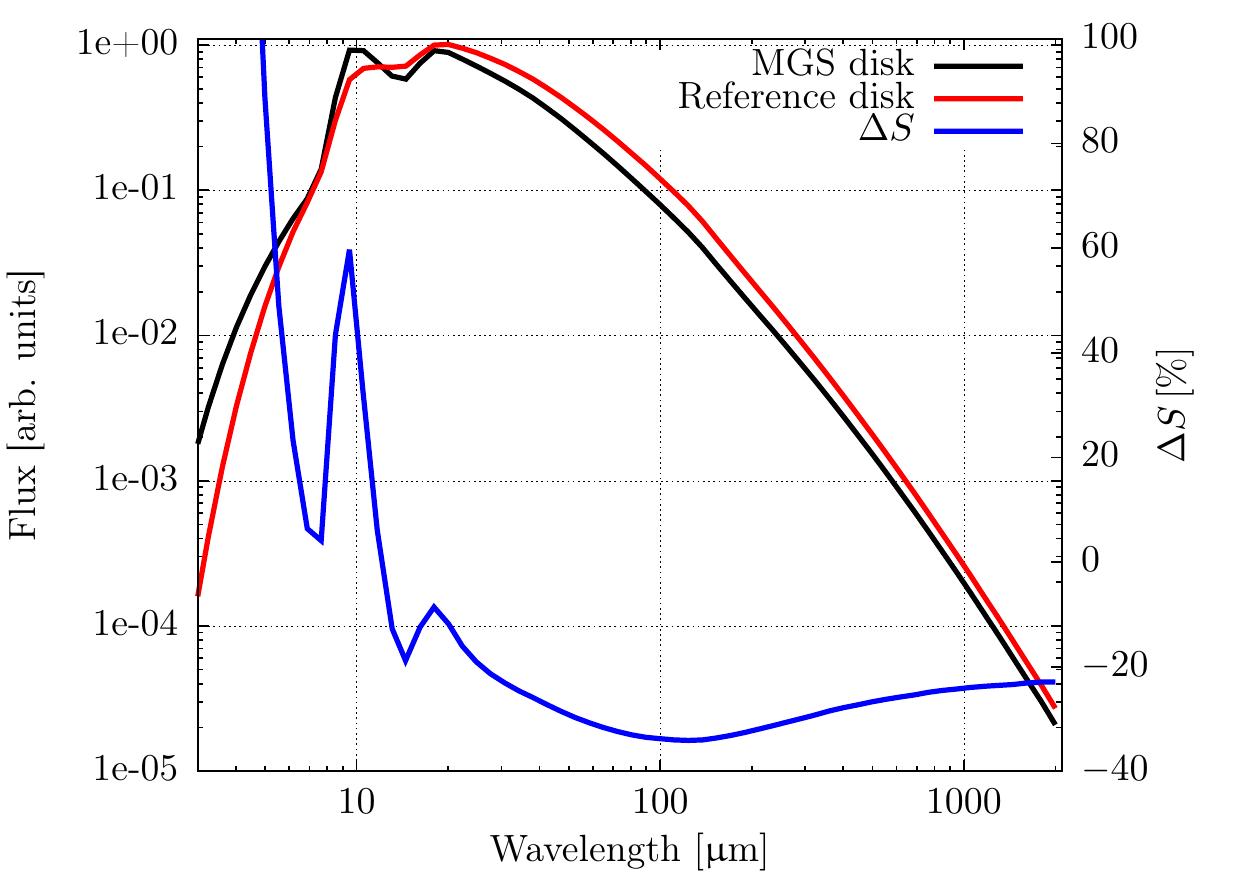}
  \caption{Thermal emission of the disk together with the relative difference
  in the optically thin case. For details, see Sect.\,\ref{seq:rt1}.}
  \label{img:raytrace1e20}
 \end{figure}

\begin{figure}
  \centering
  \includegraphics[width=1.0\columnwidth]{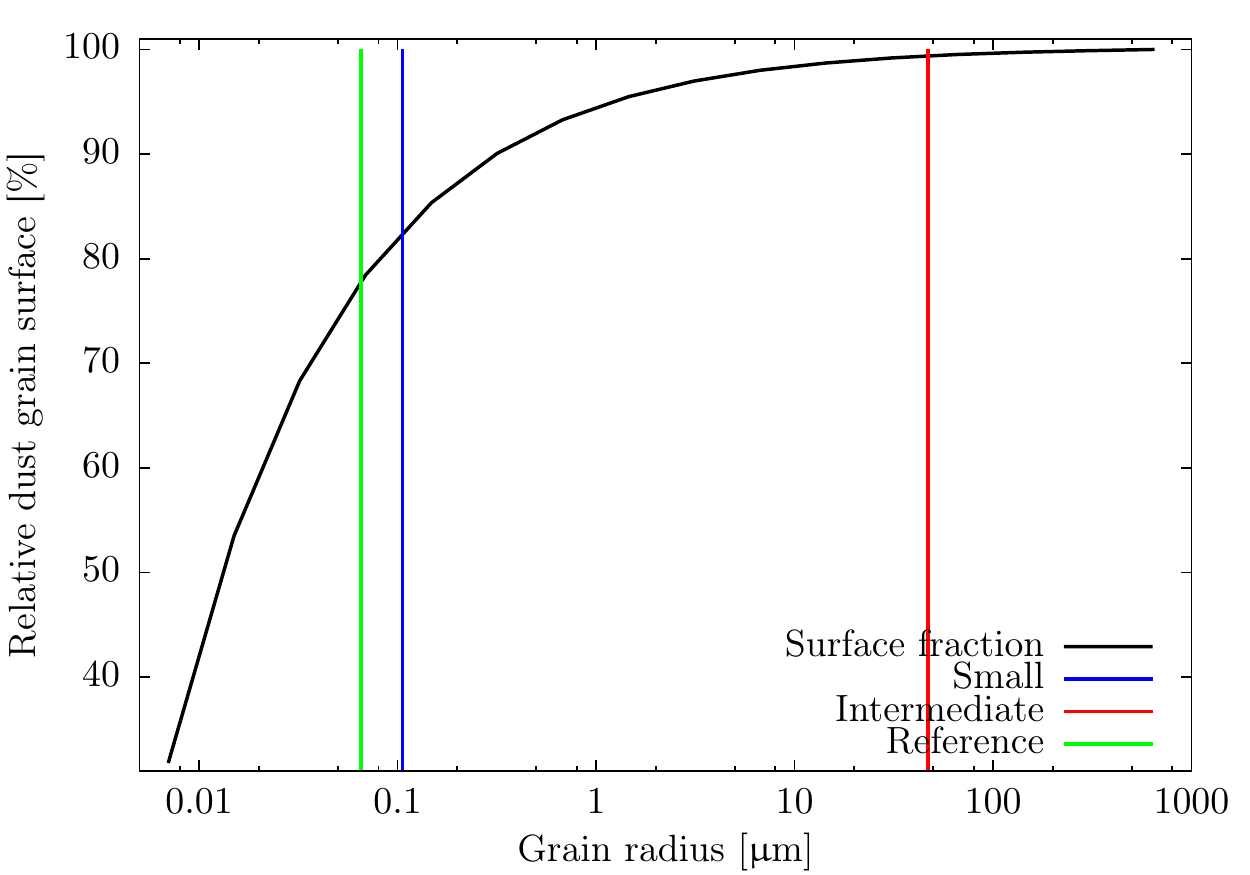}
  \caption{Relative dust grain surface of all grains below a certain radius, 
plotted cumulatively.
  In addition, the upper limits for the small-
  (radii of $5\,\mathrm{\nano \metre}$ to $106\,\mathrm{\nano \metre}$)
  and intermediate-sized grains (radii of $50\,\mathrm{\nano \metre}$ to 
$47\,\mathrm{\micro \metre}$)
  are plotted.
  The upper limit for the largest grains (radii of $22\,\mathrm{\micro \metre}$ 
to $1\,\mathrm{\milli \metre}$)
  is identical to the right boundary of the diagram. 
  As the average and not the maximum radius of a certain
  bin is drawn, there is a gap at the right boundary of the diagram.
  For details, see Sect.\,\ref{seq:rt1}.}
  \label{img:oberfl}
 \end{figure}

 We now discuss whether there is a significant difference in the SED that 
would have an impact on the 
 analysis of disk observations.
 For this purpose, the thermal emission of the disk is shown in 
Fig.\,\ref{img:raytrace1e20},
 together with the relative difference between the cases of the MGS disk and 
the reference disk
 \begin{align}
 \Delta S = 
\frac{S\!_\mathrm{\lambda}(\text{MGS})-{S}\!_\mathrm{\lambda}(\text{Reference})}
{{S}\!_\mathrm{\lambda}(\text{Reference})}. 
 \end{align}
 The radiation of the MGS disk is found to be stronger at short wavelengths 
($\lesssim13\,\mathrm{\micro\metre}$) because of the 
influence of the small grains, which 
  contribute most to the grain surface, radiate efficiently at 
short wavelengths,
 and are warmer than the average
  grains in the reference disk.
 To illustrate the contribution of the different grains to the total dust grain 
surface, 
 the relative dust grain surface below a certain grain radius
 is shown in Fig.\,\ref{img:oberfl}. In the MGS disk, dust grains 
smaller than the average grains in the reference
 account for more than $75\,\%$ of the total dust grain surface. 
 
 However, the relative difference has a local minimum at wavelengths of $\sim 
100\,\mathrm{\micro \metre}$,
 afterward the relative difference decreases again. This is because the 
absorption cross section of the average grains
 also decreases with longer wavelengths. Thus, the influence of the large 
grains, which also have a high absorption
 cross section at long wavelengths, but a small abundance, reduces the 
relative 
difference again.
 The feature around $10\,\mathrm{\micro \metre}$ is caused by the silicate 
emission,
 which is more pronounced in the thermal emission of the MGS disk.
 Thus, the wavelength dependency of the thermal emission at long wavelengths is 
affected. 
Therefore, using a more correct grain radius
 distribution also affects the dust emissivity index.
 
\begin{figure}
  \centering
  \includegraphics[width=1.0\columnwidth]{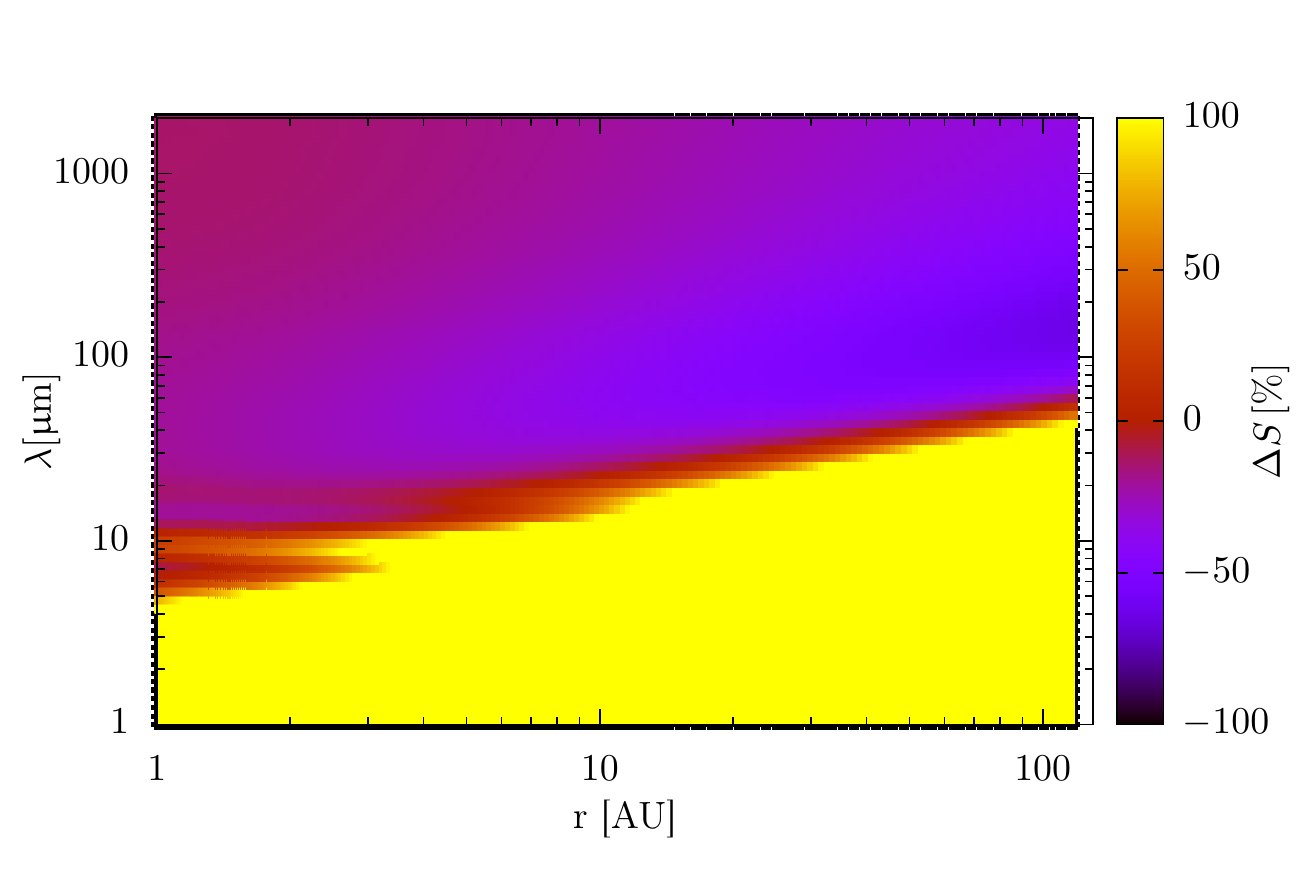}
  \caption{Relative difference in the thermal emission between the MGS disk and 
the reference disk, in the optically thin case.
  The structureless yellow region represents all cases in which the relative 
difference is $\geq1$.
  For details, see Sect.\,\ref{seq:rt1}.}
  \label{img:qb1e20}
 \end{figure}

 To further assess the differences in the thermal emission, its relative 
difference between the MGS disk
 and reference disk is computed spatially resolved throughout the disk. For 
this purpose, we compute the flux 
 $S\!_\mathrm{\lambda}(r)\sim B_\mathrm{\lambda}(T_i)\cdot 
Q_{\text{abs}\lambda}(a_i)\cdot A(a_i)$ as a function of the radial distance to 
the central star.
 Here, $r$ is the radial distance, $B_\mathrm{\lambda}$ is the Planck function, 
$Q_{\text{abs}\lambda}=C_{\text{abs}\lambda}/\pi a_\text{i}^2$ is the 
 absorption efficiency, and $A(a_i)$ is the total surface of all grains with 
radius $a_\text{i}$ in the considered volume.
 The radially resolved relative difference between the thermal emission of the 
MGS disk and the reference disk is shown in Fig.\,\ref{img:qb1e20}.
The results are similar to those presented in Fig.\,\ref{img:raytrace1e20}, but 
the position
of the minimum of the relative difference depends on the radial 
distance and
therefore on the temperature of the grains. For cooler grains further out, this 
minimum
is shifted toward longer wavelengths and getting steeper.
This is caused by the increasing relative temperature spread toward 
the outer edge.
 
 \subsection{Optically thick case}
 
 While the above results (Sects.\,\ref{seq:mtd1} to \ref{seq:rt1}) are 
applicable to the upper,
 optically thin layers of protoplanetary disks, the inner dust-depleted regions
 of transition disks as well as debris disks,
 we now explore the effect of an increase of the optical depth.
 Scattering and re-emission (see Sect.\,\ref{seq:rt2}) and also 
shielding of the midplane by
 dust above the midplane are essential in this case.
 Furthermore, owing to the resulting better thermal coupling, the temperature 
spread is much
 smaller.
 
 \subsubsection{Midplane temperature distribution}
 \label{seq:mtd2}
 
 Both the temperature further away from the star and the temperature 
 spread are much smaller than in the optically thin disk (maximum temperature 
spread $\sim 5\,\mathrm{\kelvin}$). Close to the star (within a radial 
distance of $<1.2\,\mathrm{AU}$), the midplane temperature is increased by the 
back warming of the dust in the inner disk regions.
This back warming happens at longer wavelengths than the heating by 
the star, for which the heating of the large grains is 
more efficient than for the smaller grains. This leads to a partly reversed 
order of radial temperature profiles.

\begin{figure}[!h]
  \centering
  \begin{subfigure}[b]{1.0\columnwidth}
  \includegraphics[width=1.0\columnwidth]{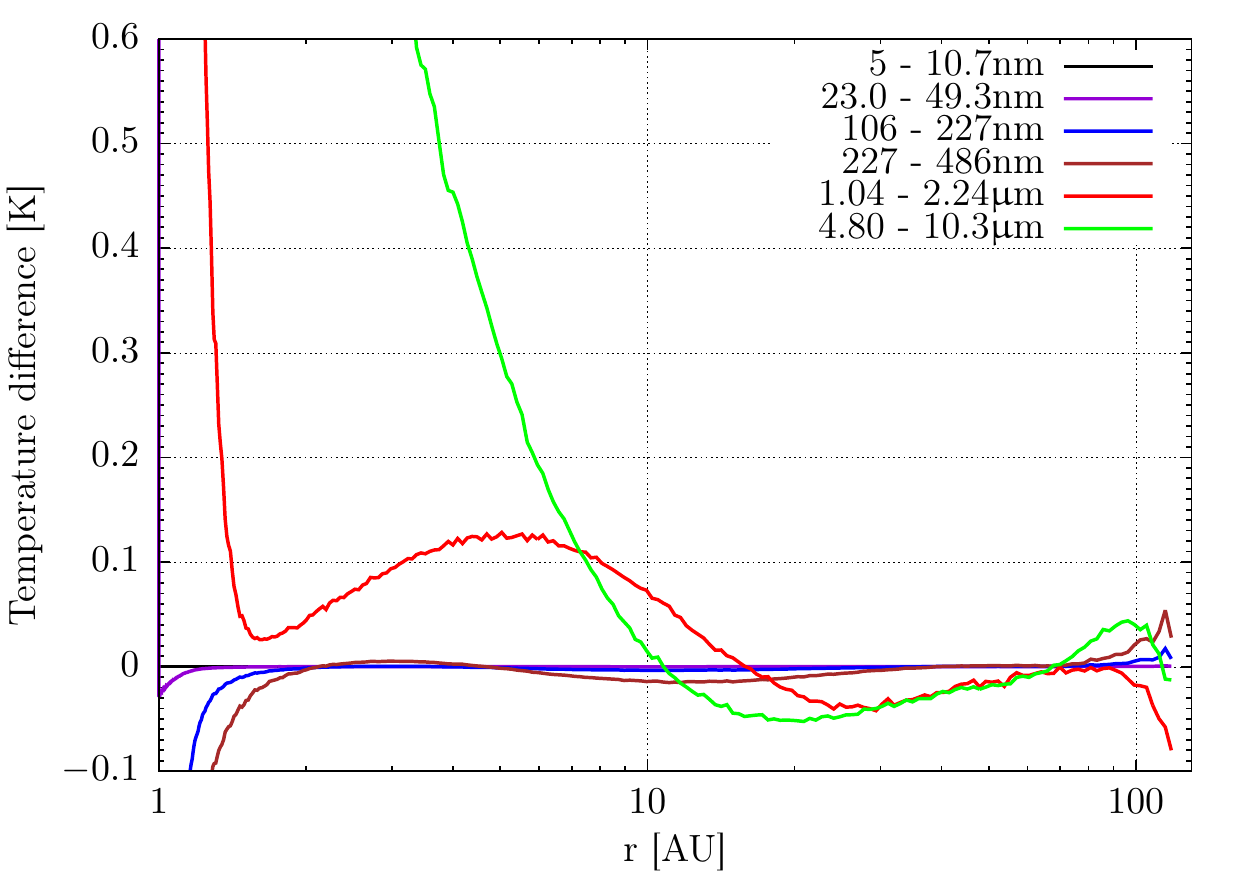}
  \end{subfigure}
\begin{subfigure}[b]{1.0\columnwidth}
\includegraphics[width=1.0\columnwidth]{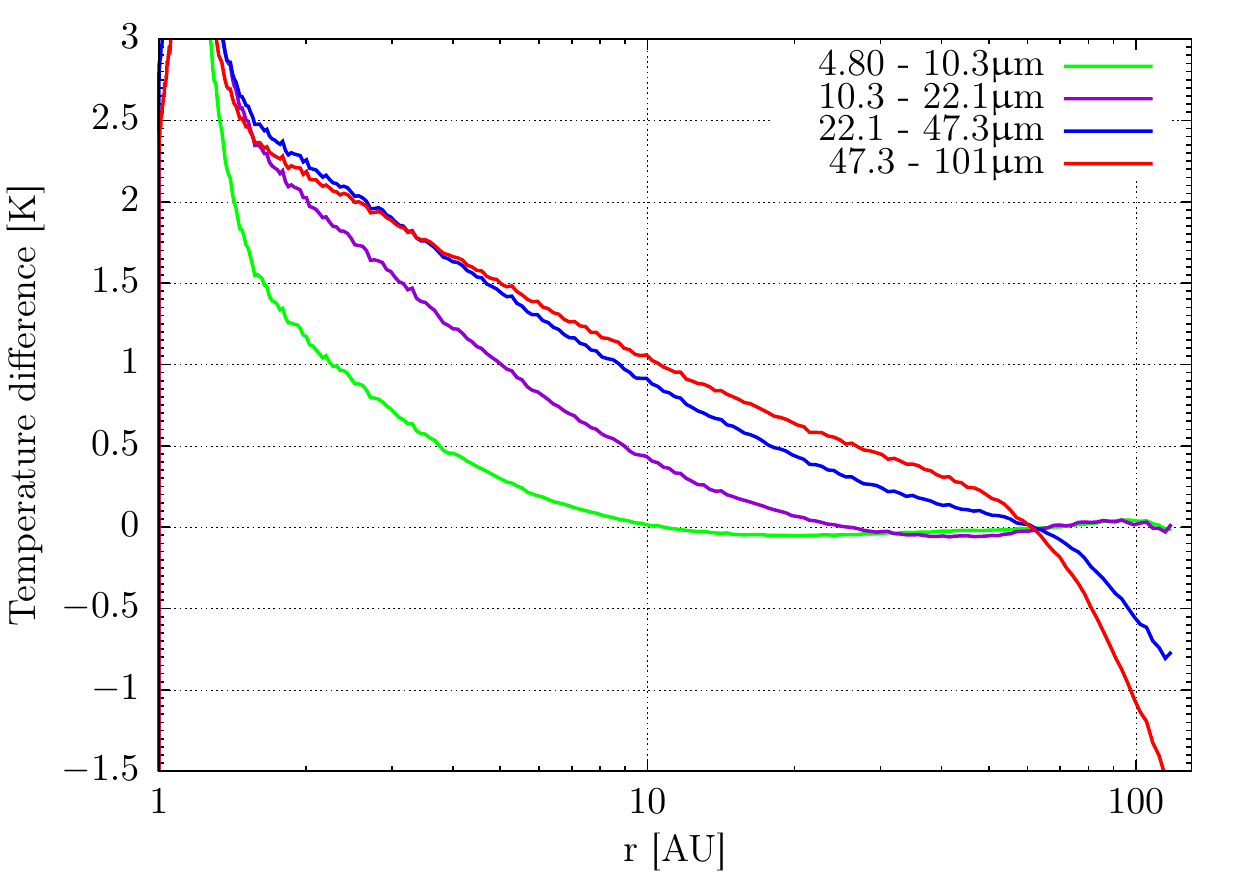}
\end{subfigure}
\begin{subfigure}[b]{1.0\columnwidth}
  \includegraphics[width=1.0\columnwidth]{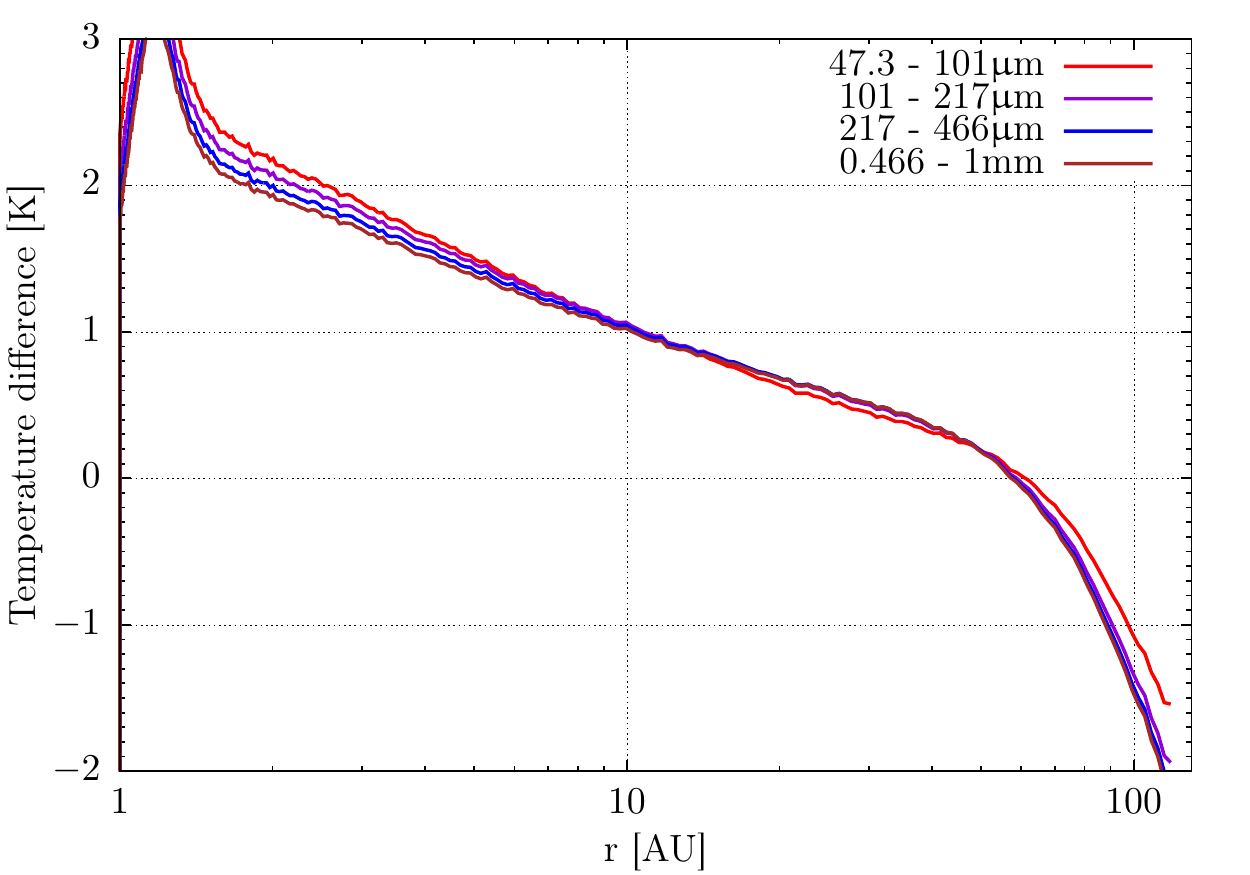}
\end{subfigure}
\caption{Differences of the midplane temperatures to the smallest dust 
grains 
(radius $5-11\,\mathrm{\nano \metre}$) in the optically thick case. For 
details, see Sect.\,\ref{seq:mtd2}.}
\label{img:diff1e3}
 \end{figure}

 Because of the small temperature spread, the difference in the 
temperature 
distribution from the smallest dust 
 grains rather than the difference from the reference disk is drawn in 
Fig.\,\ref{img:diff1e3}.
 Here, the temperature first decreases with increasing grain radii up to a
 radius between $200\,\mathrm{\nano \metre}$ and $10\,\mathrm{\micro \metre}$. 
 Then, the temperature increases with increasing grain radii up to a radius
 between $20\,\mathrm{\micro \metre}$ and $100\,\mathrm{\micro \metre}$ before 
the temperature 
 decreases again.
 
 \subsubsection{Vertical temperature structure}
 \label{seq:vts}
 
\begin{figure}[!h]
  \centering
  \begin{subfigure}[b]{1.0\columnwidth}
   \includegraphics[width=1.0\columnwidth]{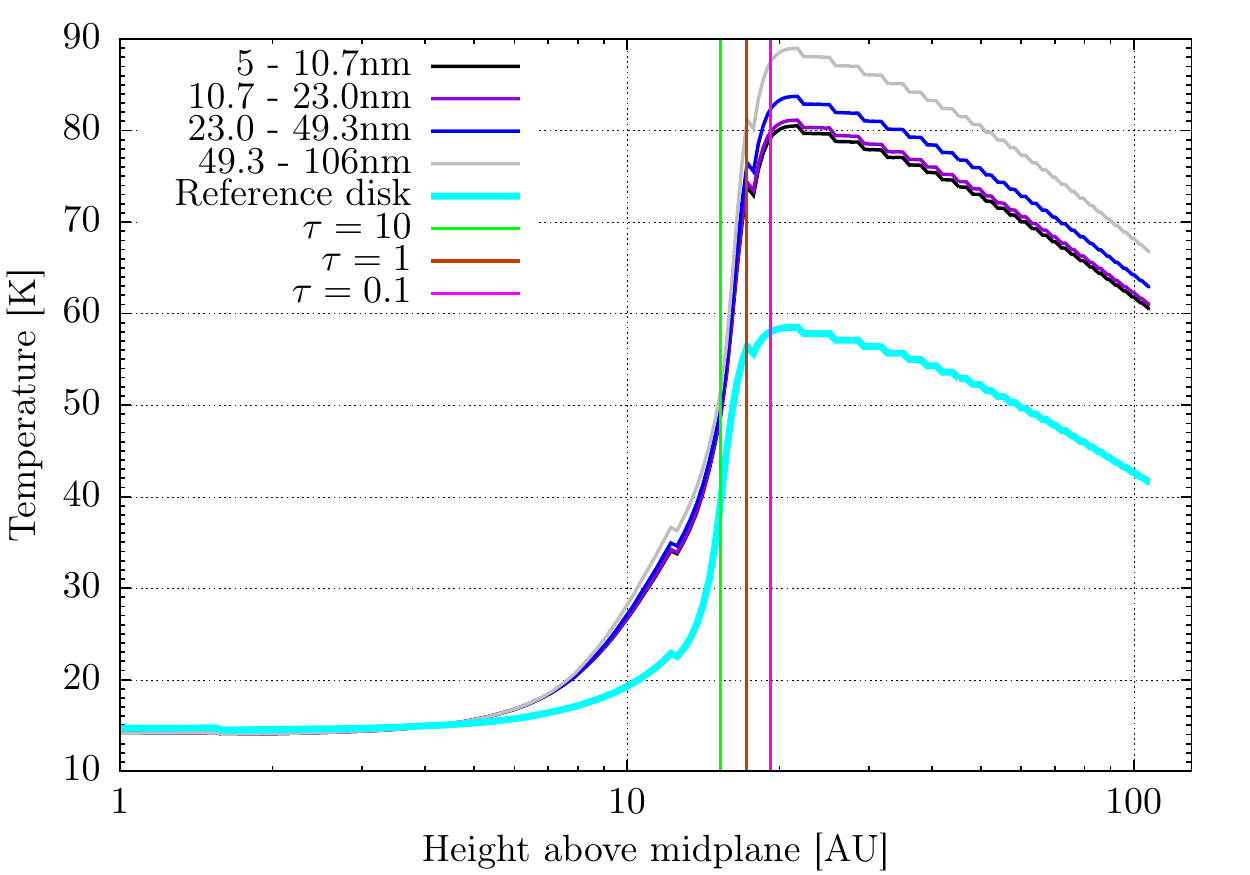}
  \end{subfigure}
\begin{subfigure}[b]{1.0\columnwidth}
\includegraphics[width=1.0\columnwidth]{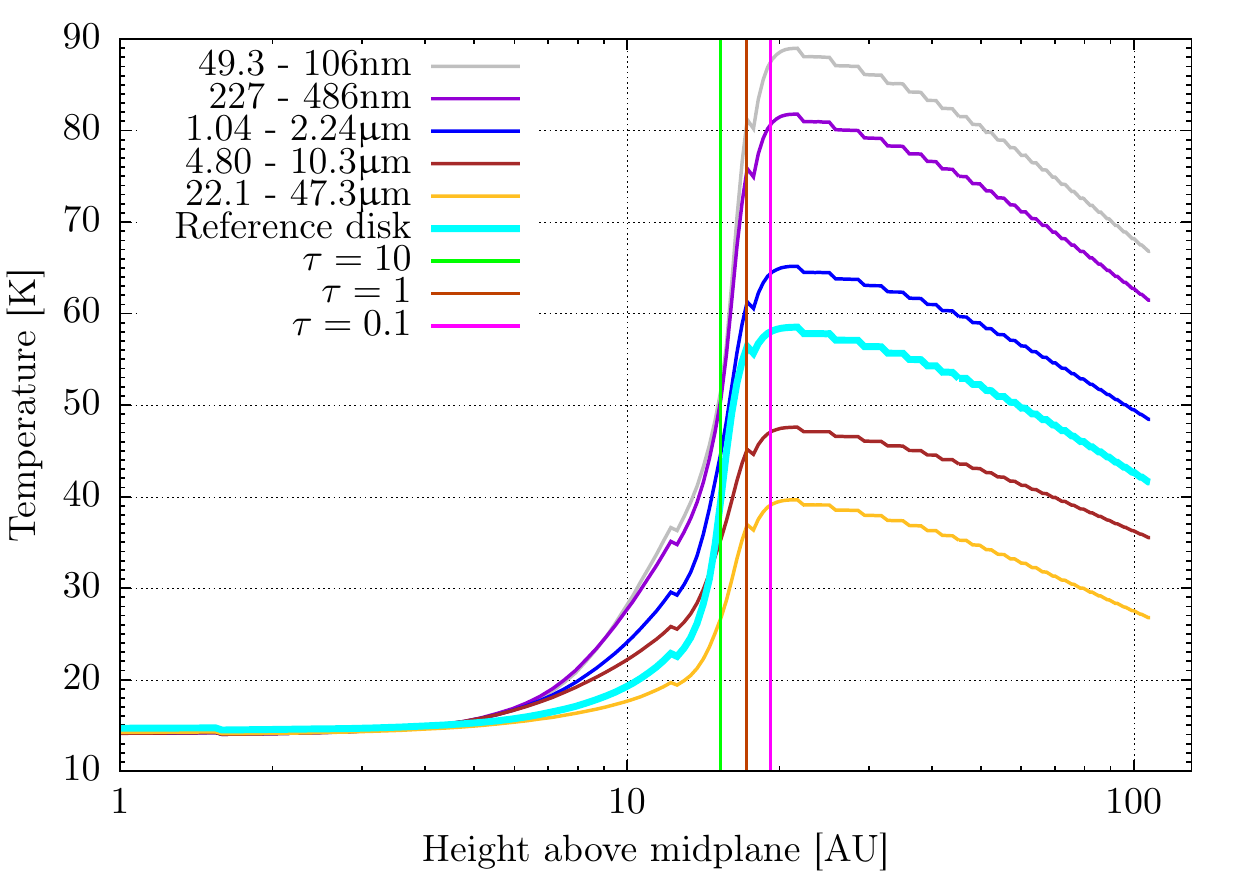}
\end{subfigure}
\begin{subfigure}[b]{1.0\columnwidth}
   \includegraphics[width=1.0\columnwidth]{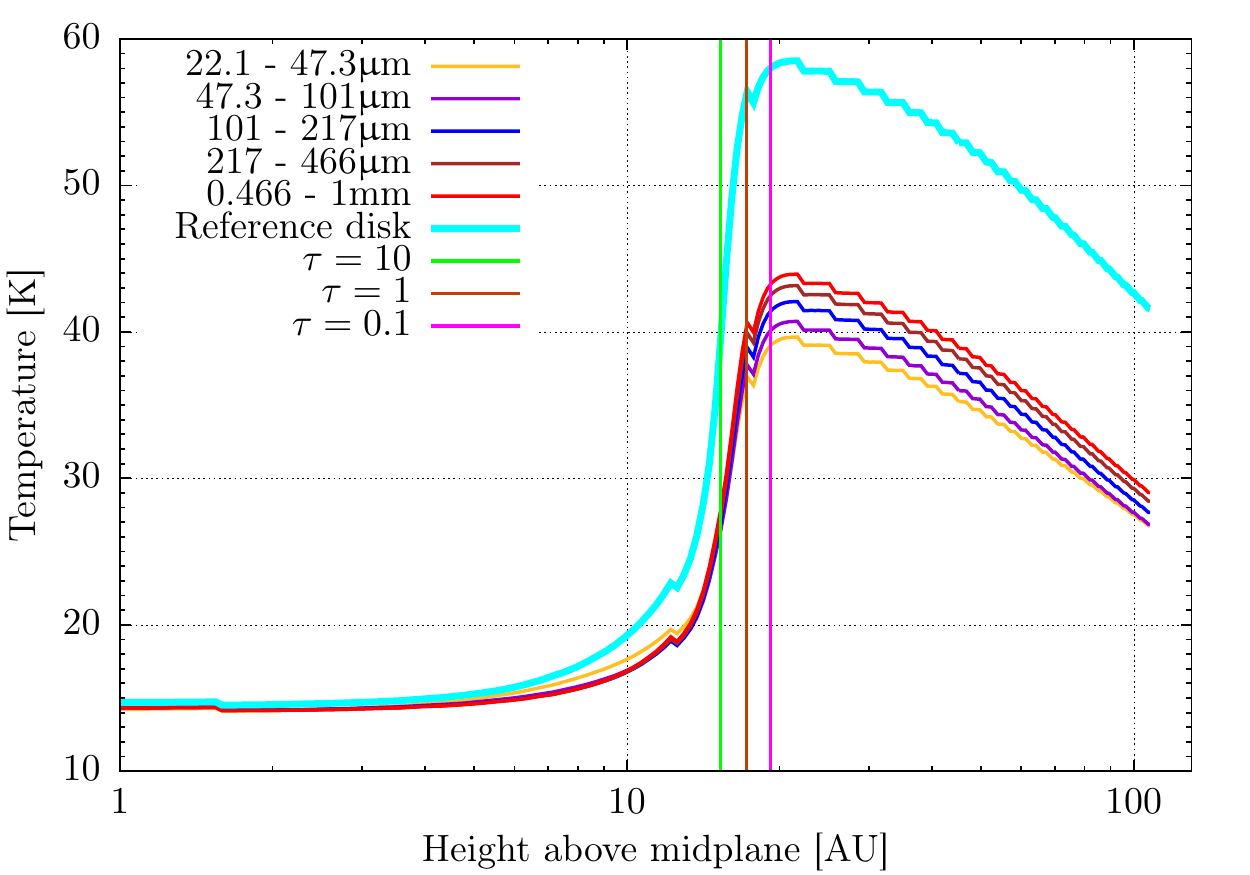}
\end{subfigure}
\caption{Vertical temperature distribution at a radial distance of
  $50\,\mathrm{AU}$ from the star in the optically thick case. For details,
  see Sect.\,\ref{seq:vts}.}
  \label{img:v1e3cutz50}
 \end{figure}

 In the upper layers of the disk, the optical depth is much smaller than in the 
midplane. Consequently,
 the temperature distribution should be similar to the optically thin case.
 To verify this assumption, the temperature distribution in 
z-direction,
 $50\,\mathrm{AU}$ from the star, is drawn in 
Fig.\,\ref{img:v1e3cutz50}.
 
  Above the $\tau = 1$ plane, the temperature first
 increases up to a grain radius of $106\,\mathrm{\nano \metre}$, then it 
decreases 
 with increasing grain radii up to a radius of $47\,\mathrm{\micro \metre}$
 before the temperature increases again.
 The spread reaches $60\%$ of the temperature value of the hottest 
grains, so 
it is a bit lower 
 than in the midplane of the optically thin disk,
 but the order of magnitude is similar.
 Qualitatively, the explanations given in Sect.\,\ref{seq:mtd1} are still valid.
 In contrast to the optically thin case, however, the grains here are not only 
heated
 by the central star, but also by the radiation of the disk. As this radiation 
is at 
 longer wavelengths, where heating is most efficient for grains with 
radii between
 $22\,\mathrm{\micro \metre}$ and $47\,\mathrm{\micro \metre}$ (which are the 
coldest grains),
 the temperature spread is slightly reduced.
 
 \subsubsection{Grain surface}
 \label{seq:gs2}
 
\begin{figure}
  \centering
  \includegraphics[width=1.0\columnwidth]{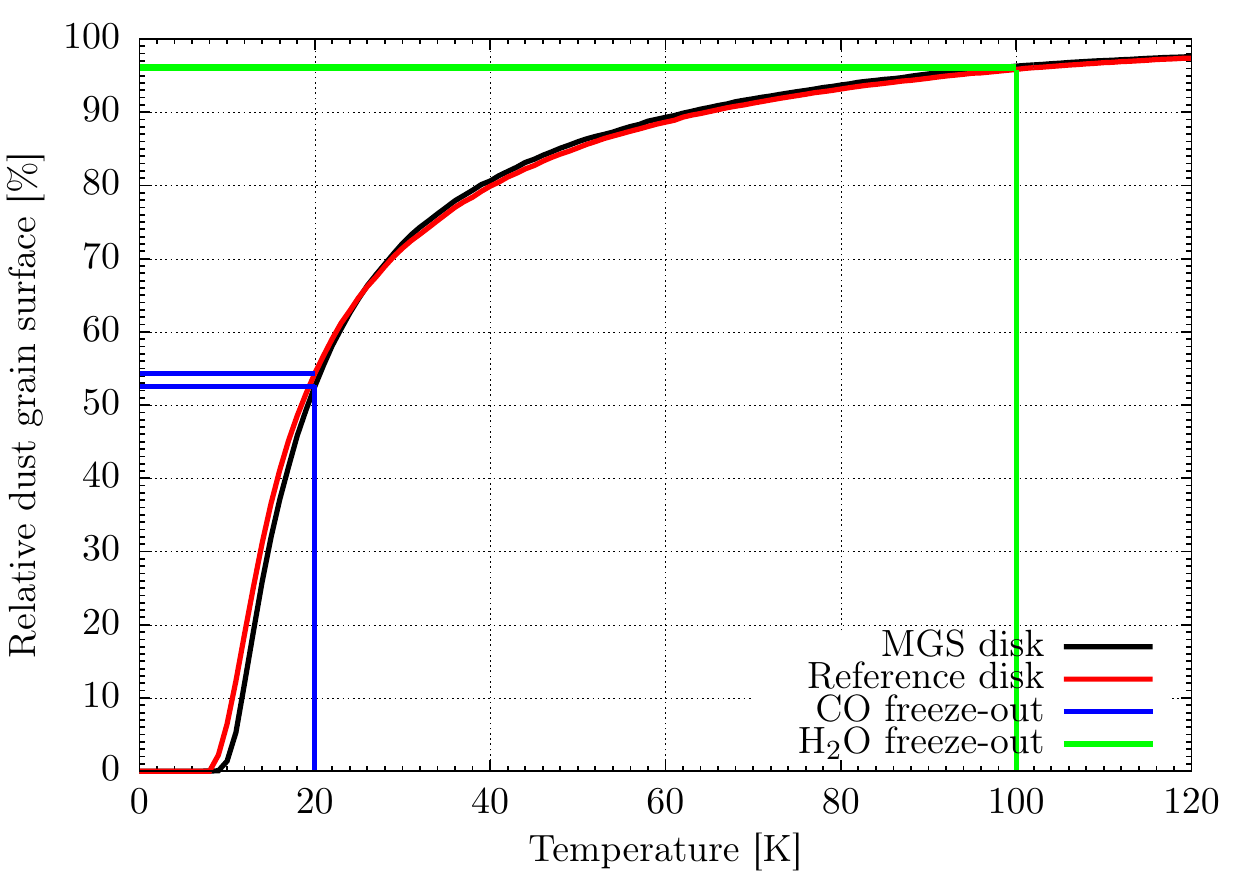}
  \caption{Relative dust grain surface below a certain temperature
  in the optically thick case. For details, see Sect.\,\ref{seq:gs2}.}
  \label{img:gs1e3}
 \end{figure}

 Although the temperature spread is much smaller, the relative dust
 grain surface below a certain temperature can still be altered and therefore 
this 
quantity needs to be
 reinvestigated (Fig.\,\ref{img:gs1e3}).
 
 We find that the dust grain surface below a given temperature is higher for 
the reference disk than for the MGS disk.
 However, for the latter it is growing faster, so that above $\sim 
30\,\mathrm{\kelvin}$, the fraction is higher for 
 the MGS disk. Also, the better thermal coupling decreases the 
 difference in this fraction.
 At $20\,\mathrm{\kelvin}$ (freeze-out temperature of CO; 
\citealt{2010ApJ...716..825O}), the relative fraction for the reference disk 
amounts to 
 $54.3\%$, while it amounts to $52.5\%$ for the MGS disk, and at 
$100\,\mathrm{\kelvin}$
 (freeze-out temperature of water; \citealt{2010ApJ...716..825O}), it amounts 
to $95.9\%$ for the reference disk and $96.3\%$ for the MGS disk.
 Between $20\,\mathrm{\kelvin}$ and $30\,\mathrm{\kelvin}$, the relative dust 
grain surface amounts
 to $17.1\,\%$ for the reference disk and $19.5\,\%$ for the MGS disk.
 
\begin{figure}
  \centering
  \begin{subfigure}[b]{1.0\columnwidth}
   \includegraphics[width=1.0\columnwidth]{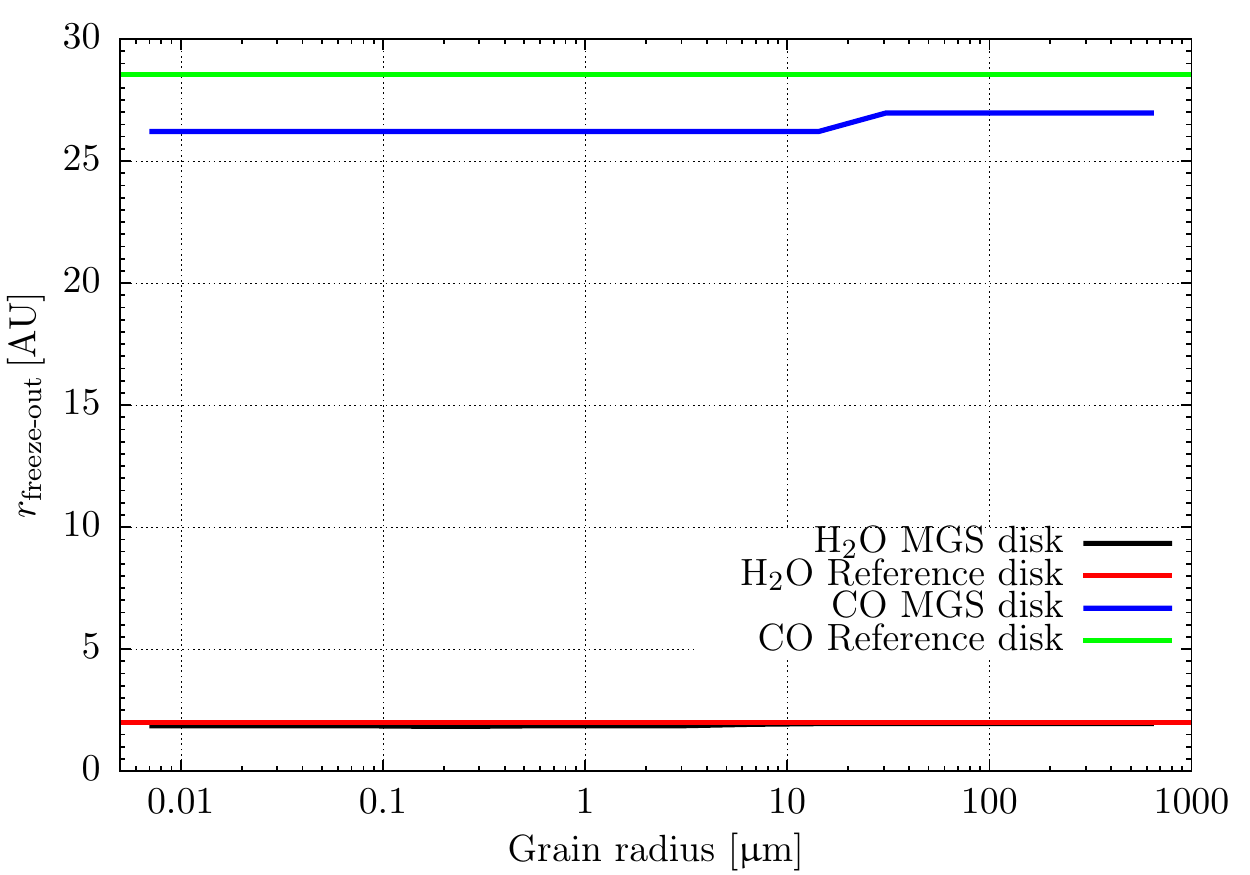}
  \end{subfigure}
\begin{subfigure}[b]{1.0\columnwidth}
 \includegraphics[width=1.0\columnwidth]{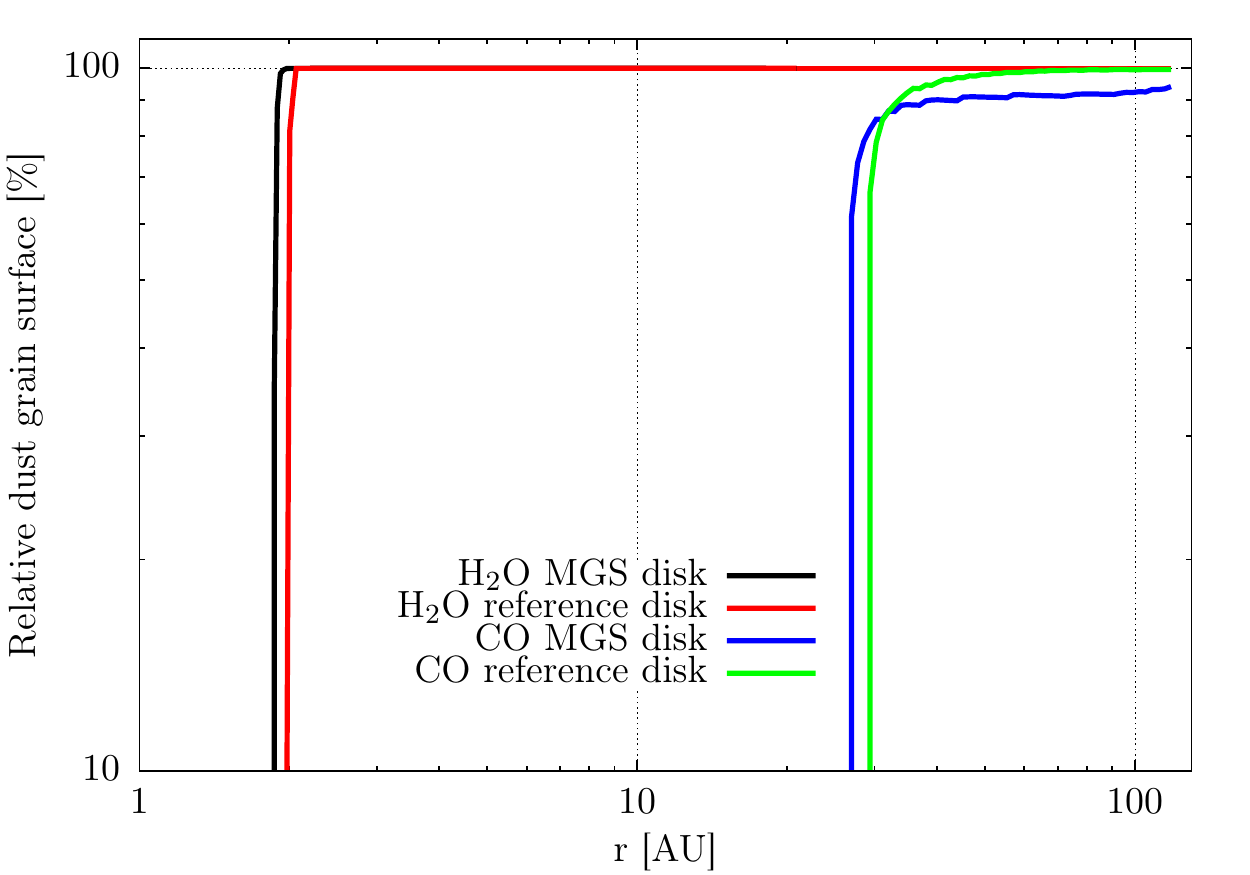}
\end{subfigure}
\caption{Top: Radial distance, outside of which the grains in 
the disk midplane are below the 
  freeze-out temperature of water ($100\,\mathrm{\kelvin}$)
  and CO ($20\,\mathrm{\kelvin}$), as a function of grain radius 
  (optically thick case). \\
  Bottom: Relative dust grain surface below the freeze-out temperature 
of 
water
  and CO, as a function of radial distance from the 
star (optically thick case).
For details, see 
Sect.\,\ref{seq:gs2}.}
\label{img:freezeout2}
 \end{figure}

 As the temperature distribution and relative dust grain surface are altered, 
the two related properties, the freeze-out radius 
(Fig.\,\ref{img:freezeout2}, top)
 and the relative dust grain surface
 below the freeze-out temperature of CO and water 
(Fig.\,\ref{img:freezeout2}, bottom), also 
need to be reanalyzed.
 
 Owing to better thermal coupling, the freeze-out radii for a volatile 
species on grains of different radii are almost the same.
For the largest grains alone (above $10\,\mathrm{\micro \metre}$), the 
freeze-out radius is larger than for the 
 smaller grains. Also, because the midplane temperature is smaller 
outside of $\sim1.2\,\mathrm{AU}$ than in the 
optically thin case, the water snowline moves further inside and 
even CO can freeze out. In addition, the 
freeze-out radius of the MGS disk is
 smaller than that of the reference disk, indicating a smaller midplane 
temperature.

 The relative dust grain surface below or equal to the freeze-out temperature 
of water ($100\,\mathrm{\kelvin}$)
 and CO ($20\,\mathrm{\kelvin}$) shows that outside of $\sim 2\,\mathrm{AU}$
 the grains below the freeze-out temperature of water constitute almost all the 
surface at that radius. CO cannot freeze out inside
 $\sim30\,\mathrm{AU}$.
 There are three reasons for the differences between the optically 
thick and the optically 
thin case 
(Fig.\,\ref{img:freezeout1}, bottom): 
 \begin{enumerate}
  \item The midplane temperature spread is much smaller, so that the fraction 
of surface
 below the freeze-out temperature increases faster with radial distance from 
the star for the MGS disk than in the optically thin case.
 \item The 
 upper layers of the disk are optically thin and warmer than the midplane, 
resulting in a smoother transition both for 
 the MGS disk and reference disk.
 \item The midplane is colder outside of $\sim1.2\,\mathrm{AU}$ than 
in the optically thin case.
 \end{enumerate}
 
 \subsubsection{Thermal emission of the disk}
 \label{seq:rt2}
 
\begin{figure}
  \centering
  \includegraphics[width=1.0\columnwidth]{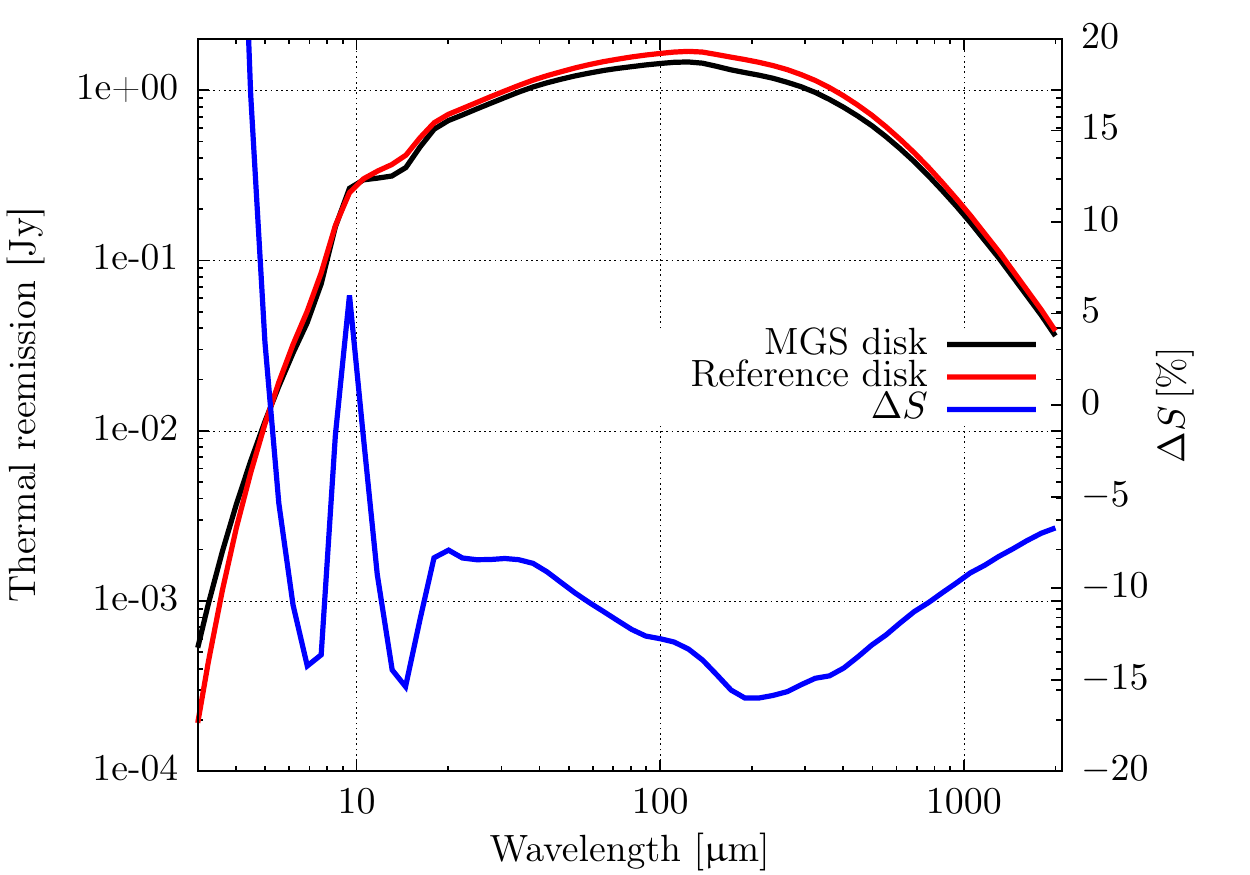}
  \caption{Thermal emission of the MGS disk and the reference disk,
  together with the relative difference of this quantity, in the optically
  thick case.
  For details, see Sect.\,\ref{seq:rt2}.}
  \label{img:raytrace1e3}
 \end{figure}

 To illustrate how the thermal emission differs from the optically thin case 
and between the two disks,
 the relative difference 
 $\Delta S = 
(S\!_\mathrm{\lambda}(\text{MGS})-{S}\!_\mathrm{\lambda}(\text{Reference}))
 /{S}\!_\mathrm{\lambda}(\text{Reference})$ is drawn
 together with the absolute flux values in Fig.\,\ref{img:raytrace1e3}.
 This plot shows qualitatively the same results as in the optically thin case.
 However, the maximum of the thermal emission at longer wavelengths is due to 
the 
 smaller midplane temperature outside the innermost region. Also, the 
relative difference is smaller because 
the 
temperature difference in the disk midplane is smaller.

 \begin{figure}
  \centering
  \includegraphics[width=1.0\columnwidth]{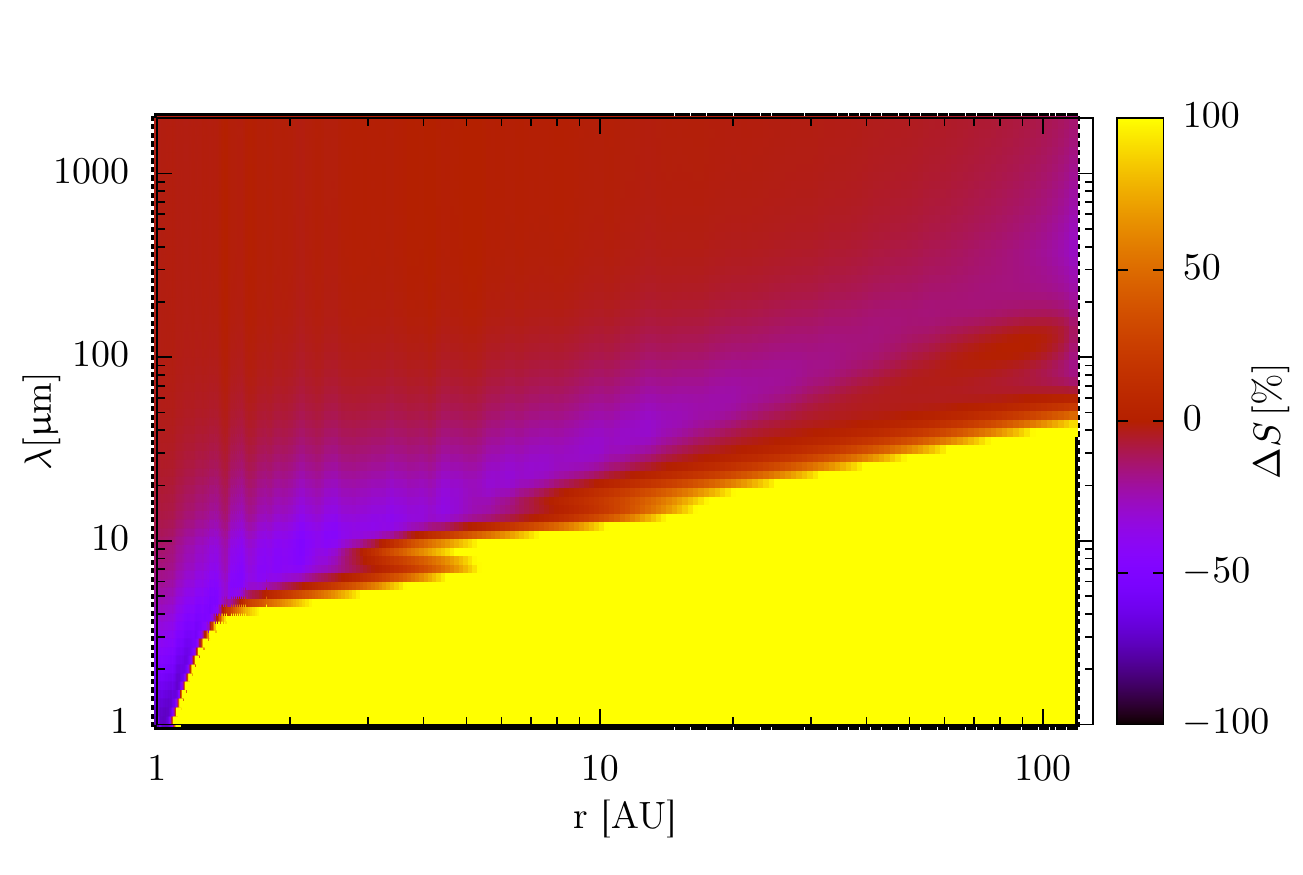}
  \caption{Relative difference in the thermal emission between the MGS disk and 
reference disk in the optically thick case. 
  The structureless yellow region represents all cases in which the relative 
difference is $\geq1$.
  For details, see Sect.\,\ref{seq:rt2}.}
  \label{img:qb1e3}
 \end{figure}
 To further investigate the differences in thermal emission, 
 the spatially resolved relative difference is shown in
 Fig.\,\ref{img:qb1e3}. As in the case of the optically thin disk 
(Fig.\,\ref{img:qb1e20}), the minimum in the relative difference moves to 
longer wavelengths with increasing radial distance. However, because of the 
reducing temperature spread the minimum is getting shallower.

 \subsection{Influence of different disk masses}
 \label{seq:masses}
 
 To emphasize the impact of the different treatments of dust
 grain radius distributions on the temperature distribution and thermal
 re-emitted radiation of a circumstellar disk, in the preceding sections we 
considered two extreme cases: 
 an optically thin (i.e., very low mass) disk and a rather massive
 disk.
 To evaluate the consequences of more realistic masses 
 on the results, we calculate the temperature 
 distribution for disks with dust masses between $10^{-5}\,\mathrm{M_{\odot}}$
 and $3\cdot 10^{-4}\,\mathrm{M_{\odot}}$.
 
\begin{figure}
  \centering
  \includegraphics[width=1.0\columnwidth]{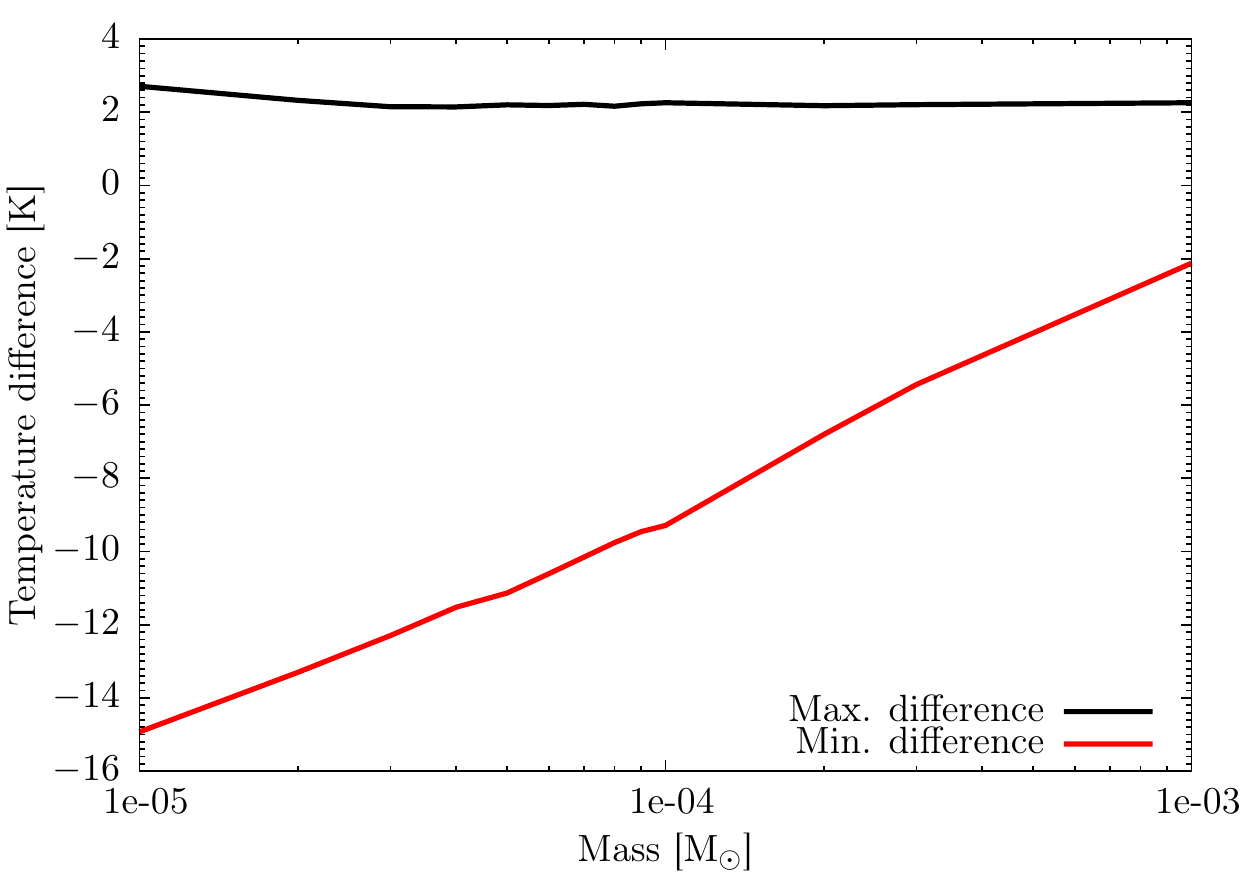}
  \caption{Trends of the maximum and minimum difference from the 
smallest dust 
grains 
  (radius $5-11\,\mathrm{\nano \metre}$) in the disk midplane. For details, 
see 
Sect.\,\ref{seq:masses}.}
  \label{img:masses}
 \end{figure}

 We find that while the results are similar to those for the optically thick 
case,
 the temperature spread increases with decreasing disk mass. This is due to the 
weaker thermal coupling.
A plot of the maximum and minimum midplane temperature difference of the 
different dust 
species to the dust species 
with the smallest grains (radius $5-11\,\mathrm{\nano \metre}$) is shown in 
Fig.\,\ref{img:masses}. For the maximum and minimum spread, the 
cells with a 
radial distance of more than $1\,\mathrm{AU}$ to the inner edge of the disk are 
considered.
The maximum temperature difference only changes by $\sim0.5\,\mathrm{\kelvin}$,
but the minimum difference changes by $\sim13\,\mathrm{\kelvin}$ from the disk 
with 
the lowest to the highest mass considered.

\begin{figure}
  \centering
  \begin{subfigure}[b]{1.0\columnwidth}
  \includegraphics[width=1.0\columnwidth]{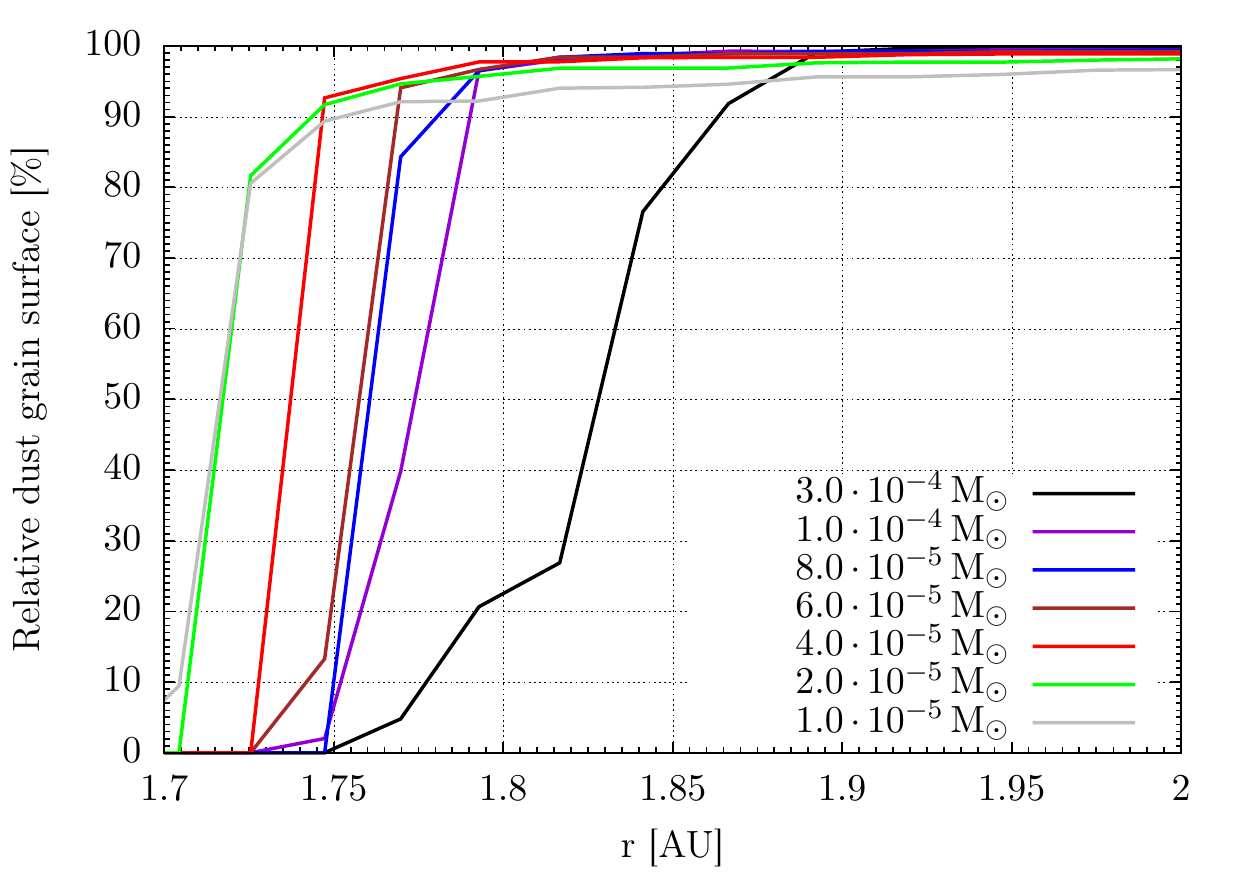}

  \end{subfigure}
\begin{subfigure}[b]{1.0\columnwidth}
\includegraphics[width=1.0\columnwidth]{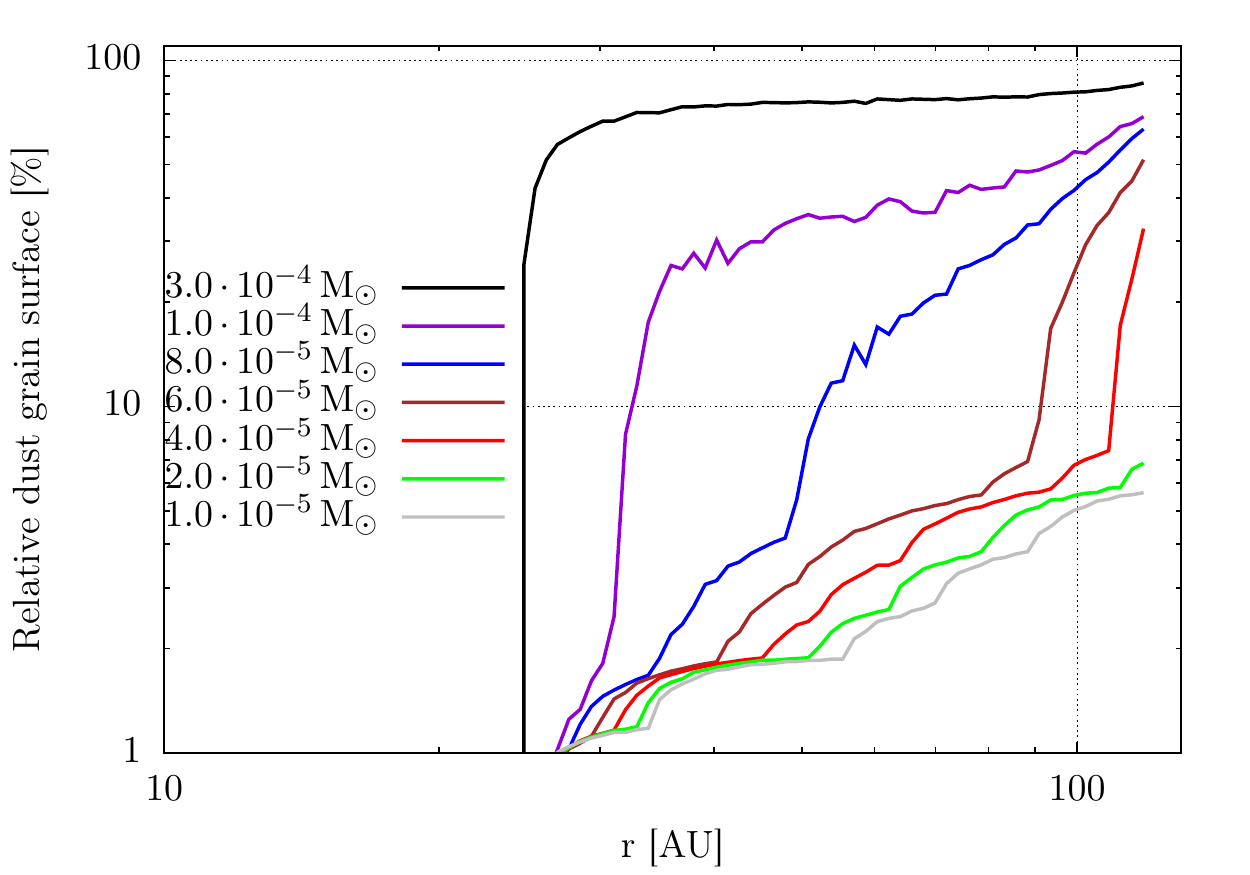}
\end{subfigure}
\caption{Relative dust grain surface below the freeze-out temperature of water 
(top) and CO (bottom) as a function of radial distance from 
the star for different 
disk masses. For details, see 
Sect.\,\ref{seq:masses}.}
\label{img:xfrmassesw}
 \end{figure}

 A different perspective on this problem is to investigate the relative dust 
grain surface with a temperature below or equal to the freeze-out 
temperature for all grains at a given radial distance from the central star 
(see Fig.\,\ref{img:xfrmassesw}). With 
decreasing mass, the radial 
distance where the freeze-out of CO happens increases due to the 
higher midplane 
temperature. Also, the increase of the dust grain surface below the 
freeze-out 
temperature of CO becomes shallower, i.e., the transition region from 
no freeze-out 
to complete freeze-out is spread over an increasingly large radial range.
The dust grain surface below the freeze-out temperature of water shows 
the opposite trend. This is due to the higher midplane temperature of higher 
mass disks. This effect can be explained by the back warming in the inner disk 
regions, which increases with increasing optical depth or disk mass. However, 
at 
the same time the temperature decrease is steeper 
for higher mass disks, so that outside of $\sim10\,\mathrm{AU}$ the 
midplane 
temperature is higher for lower mass disks.

 \subsection{Inner cavity}
 \label{seq:ic}
 
 In this section, we explore the impact of the inner inner disk radius on the 
resulting temperature spread.
 In addition to disks with an inner radius determined close to the 
sublimation temperature of the dust we now consider inner cavities with 
sizes between $5\,\mathrm{AU}$ and $50\,\mathrm{AU}$
 (outer radius $120\,\mathrm{AU}$). The assumed disk 
mass is scaled to keep the density in the remaining disk unchanged 
corresponding 
to a dust mass of $3\cdot10^{-4}\,\mathrm{M_{\odot}}$ extending from 
$1\,\mathrm{AU}$ to $120\,\mathrm{AU}$. 

\begin{figure}
  \centering
  \includegraphics[width=1.0\columnwidth]{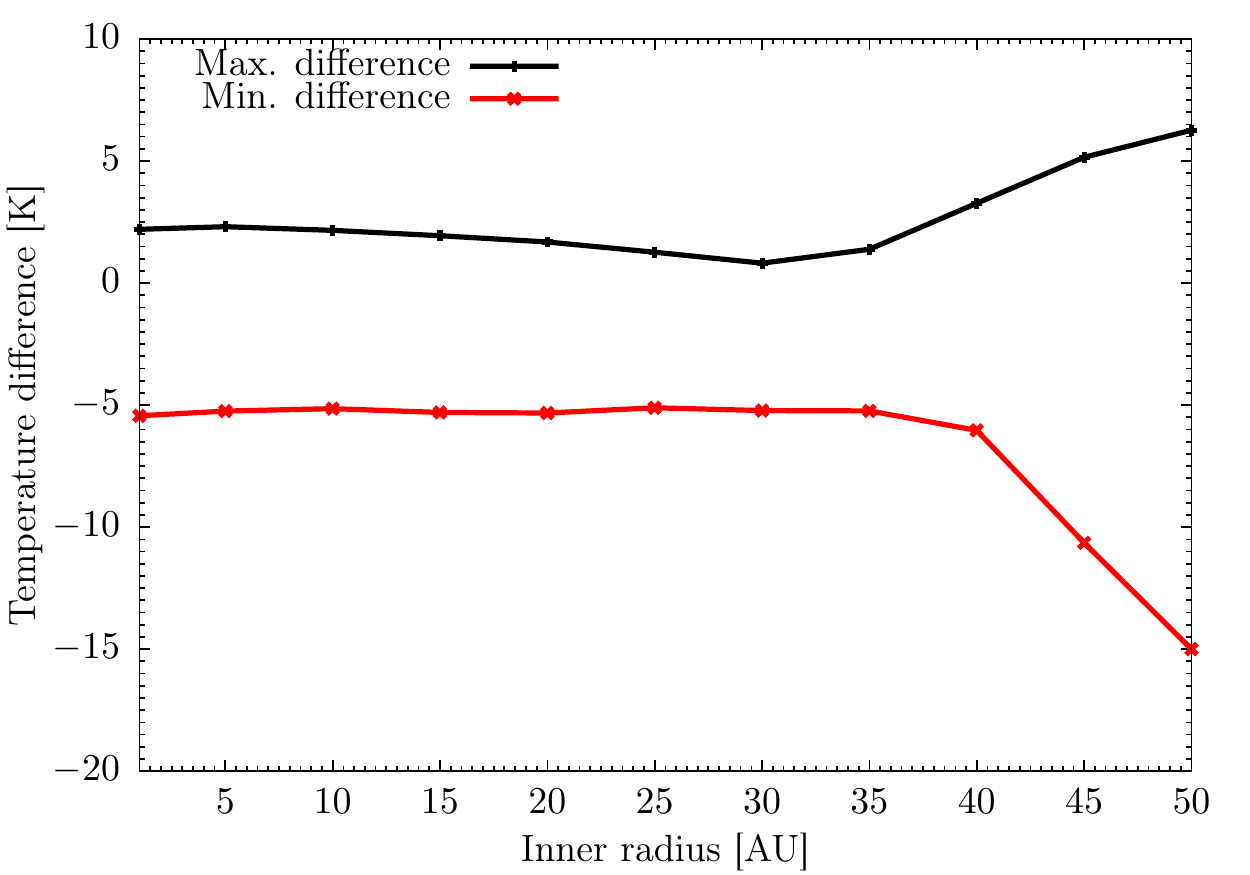}
  \caption{Trends of the maximum and minimum difference from the 
smallest dust 
grains 
  (radius $5-11\,\mathrm{\nano \metre}$) in the disk midplane. For details, 
see 
Sect.\,\ref{seq:ic}.}
  \label{img:ic}
 \end{figure}

The maximum and minimum temperature difference of the different 
grain species to the smallest dust grains 
(radius $5-11\,\mathrm{\nano \metre}$) is shown in Fig.\,\ref{img:ic}. The 
maximum temperature difference has a minimum at $30\,\mathrm{AU}$, while the 
minimum temperature difference begins to decrease for cavity sizes of more than 
$30\,\mathrm{AU}$.
The decreasing spread is due to the weaker radiation at larger radii. This 
leads to a lower temperature and thus to a smaller temperature spread. The 
increase of the spread is caused by the decreasing density in the disk, which 
leads to a weaker shielding of the dust grains by those further in and to a 
weaker thermal coupling.
 
\begin{figure}
  \centering
  \begin{subfigure}[b]{1.0\columnwidth}
 \includegraphics[width=1.0\columnwidth]{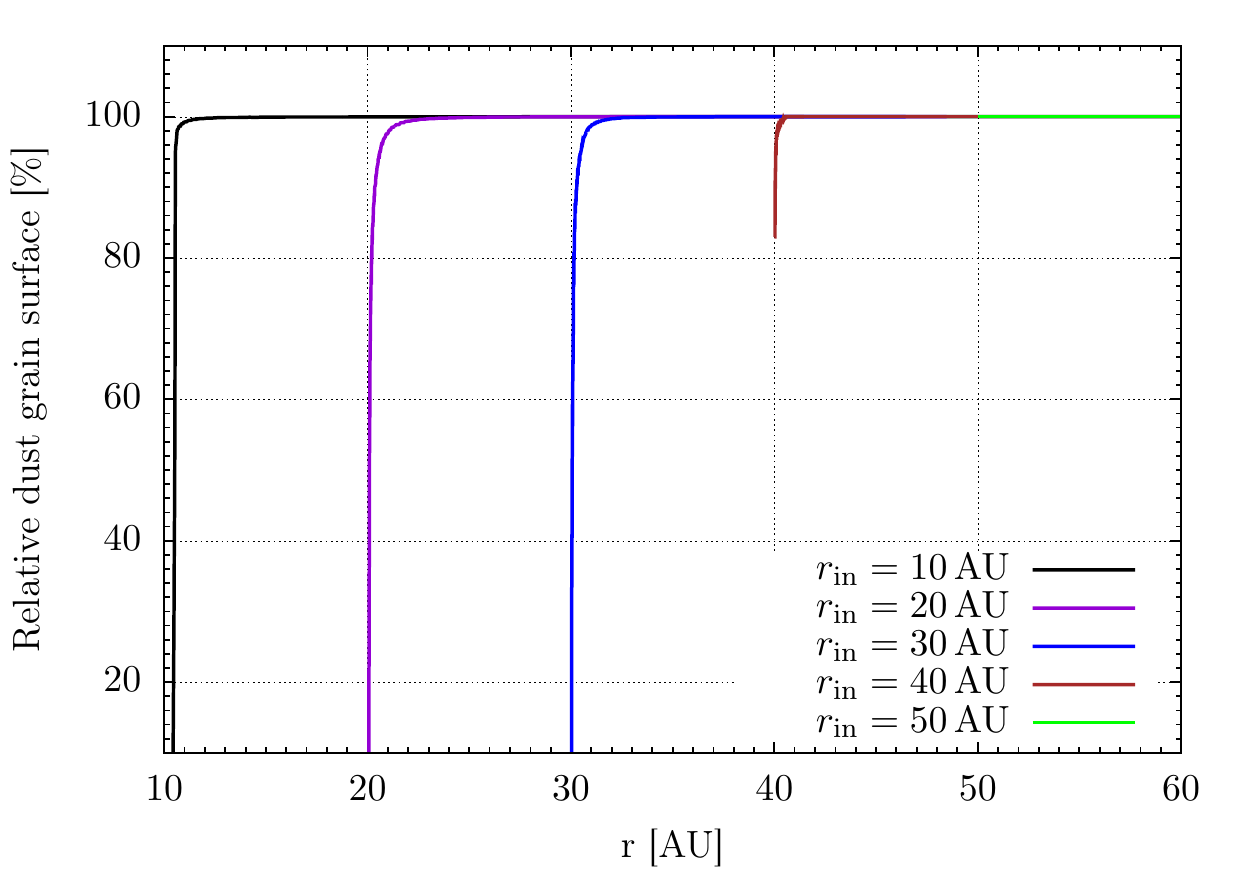}
  \end{subfigure}
\begin{subfigure}[b]{1.0\columnwidth}
\includegraphics[width=1.0\columnwidth]{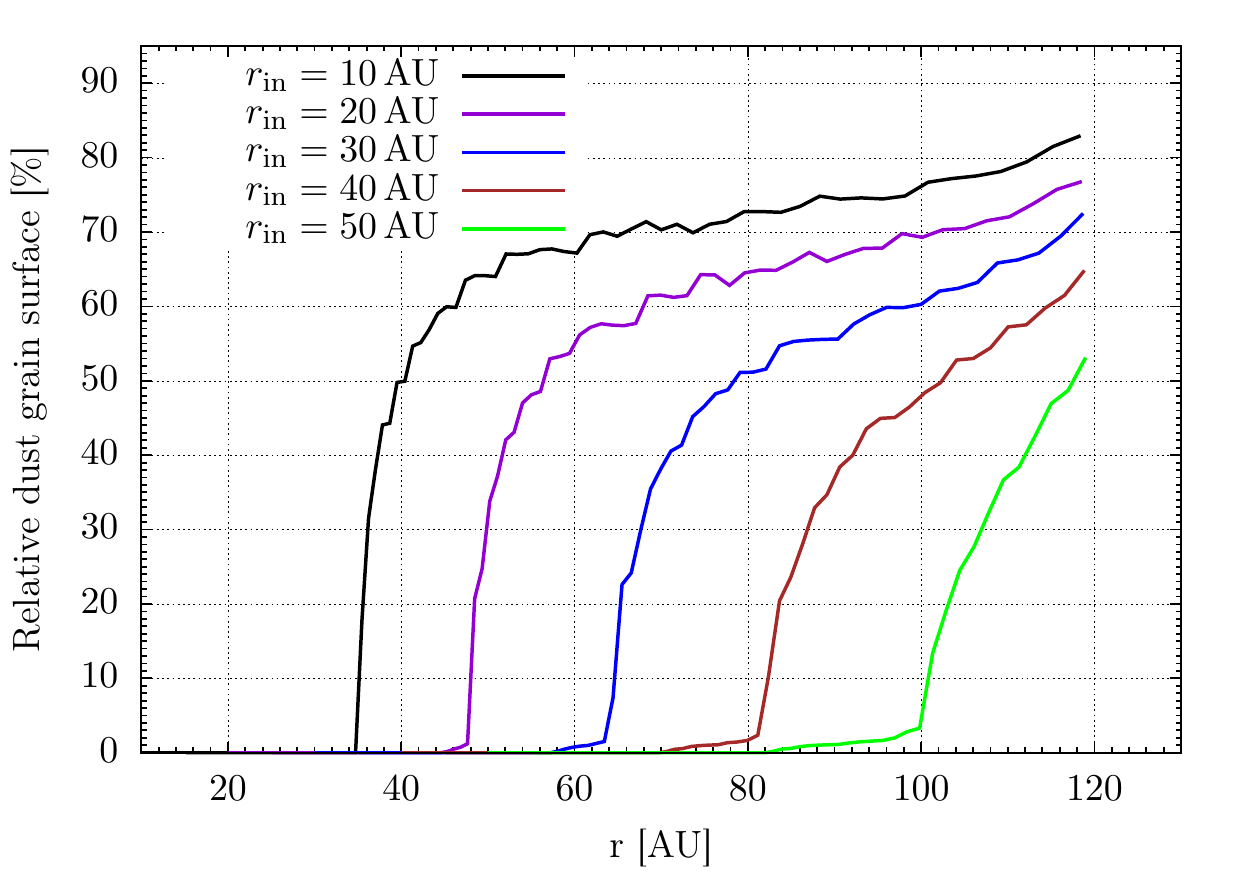}
\end{subfigure}
\caption{Relative dust grain surface below the freeze-out temperature of water 
(top) and CO (bottom) as a function of radial distance from 
the star for different 
inner radii. For details, see Sect.\,\ref{seq:ic}.}
\label{img:xfreezeoutcav}
 \end{figure}

 The relative dust grain surface 
below the freeze-out temperature of water and CO is shown in
Fig.\,\ref{img:xfreezeoutcav}. Owing to the larger inner void region, the 
stellar 
radiation can penetrate into outer disk regions more efficiently. Thus, the 
freeze-out happens at a larger radial distance for larger inner radii.
However, for the largest inner radii  
water already freezes out at the inner edge of the disk, even on the 
smallest grains.
 
 \subsection{Effects of different outer radii}
 \label{seq:or}
 
 In this chapter, we explore the effects of varying the outer radius
 of the disk from $100\,\mathrm{AU}$ to $400\,\mathrm{AU}$.
 The inner disk radius is fixed at $1.0\,\mathrm{AU}$ and the assumed disk mass 
is $3\cdot 10^{-4}\,\mathrm{M_{\odot}}$.
 
 An increase of the outer disk radius results in a decrease of the mass density 
and
 thus the optical depth of the disk. This should increase the 
 temperature spread and the freeze-out radius of CO should move 
further out, while the freeze-out radius for water should move further 
in.
However, as the dust density decreases with radial
 distance to the star, the optical depth at the inner parts of the disk
 only differs by $\sim 30\,\%$ when the outer radius of the disk is 
doubled 
(see 
Tab.\,\ref{tab:od}).
 
\begin{figure}
  \centering
  \includegraphics[width=1.0\columnwidth]{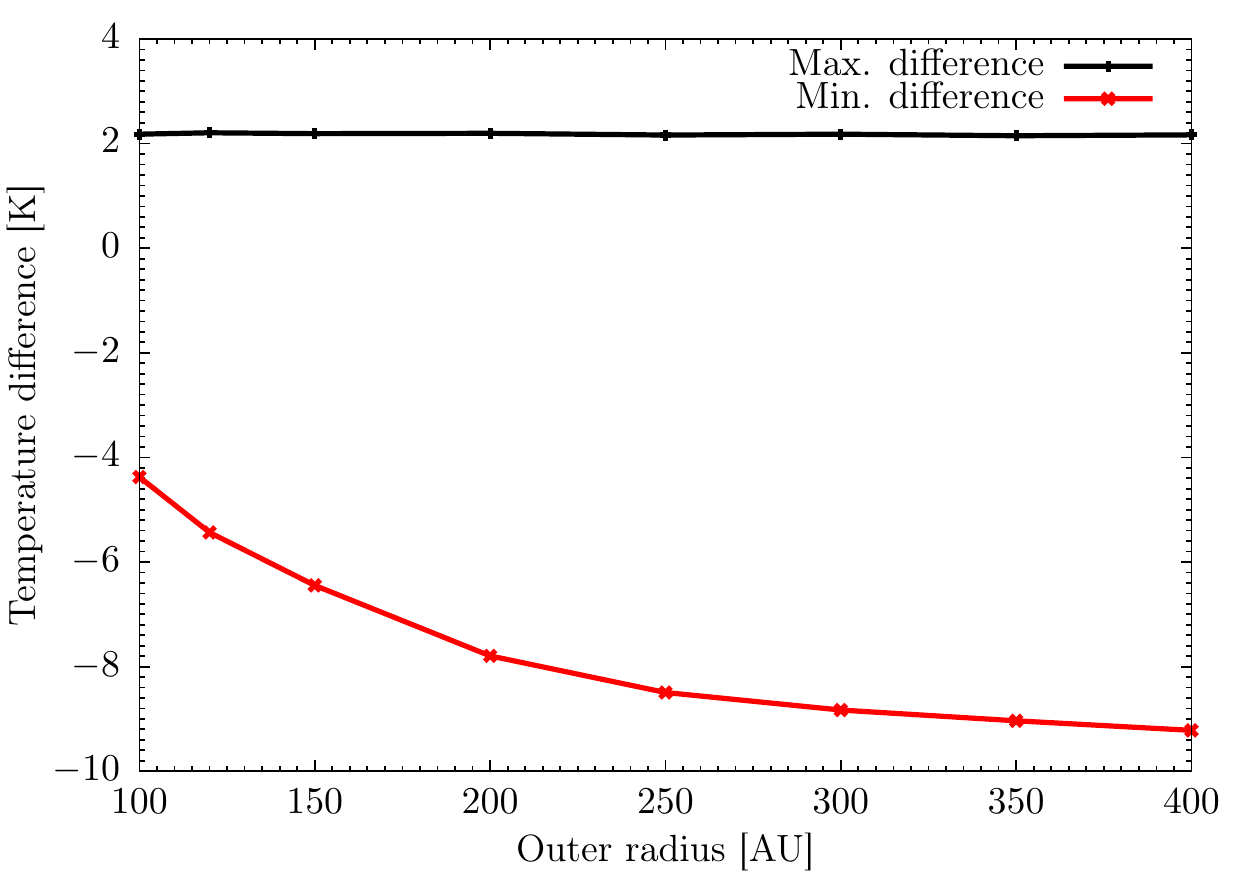}
  \caption{Trends for the maximum and minimum difference from the 
smallest dust 
grains 
  (radius $5-11\,\mathrm{\nano \metre}$) in the disk midplane. For details, 
see 
Sect.\,\ref{seq:or}.}
  \label{img:or}
 \end{figure}

 The maximum and minimum temperature difference from the smallest dust 
grains 
(radius $5-11\,\mathrm{\nano \metre}$) is shown in Fig.\,\ref{img:or}. 
Only
 the minimum temperature difference grows considerably, the maximum difference 
changes by less than $0.1\,\mathrm{\kelvin}$.
 
\begin{figure}
  \centering
  \begin{subfigure}[b]{1.0\columnwidth}
 \includegraphics[width=1.0\columnwidth]{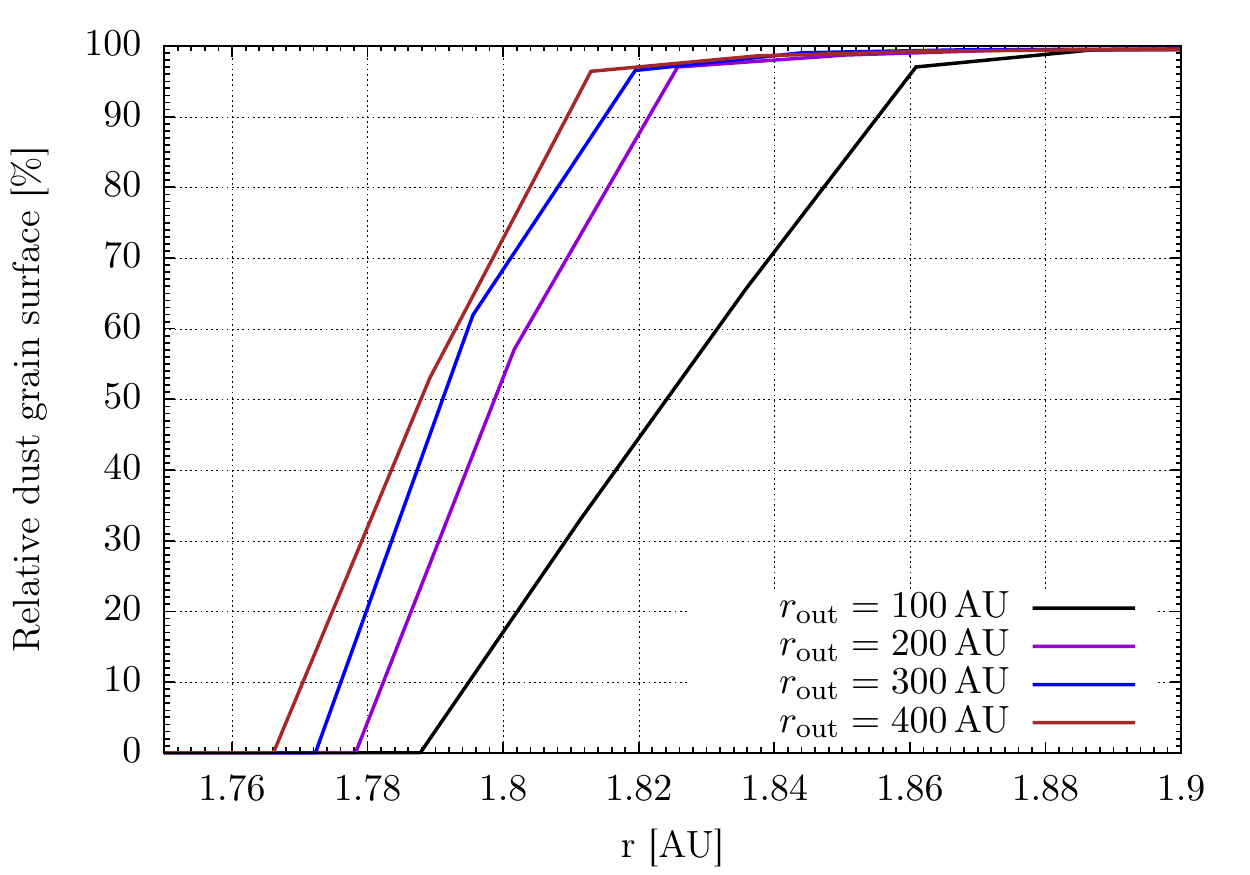}
  \end{subfigure}
\begin{subfigure}[b]{1.0\columnwidth}
\includegraphics[width=1.0\columnwidth]{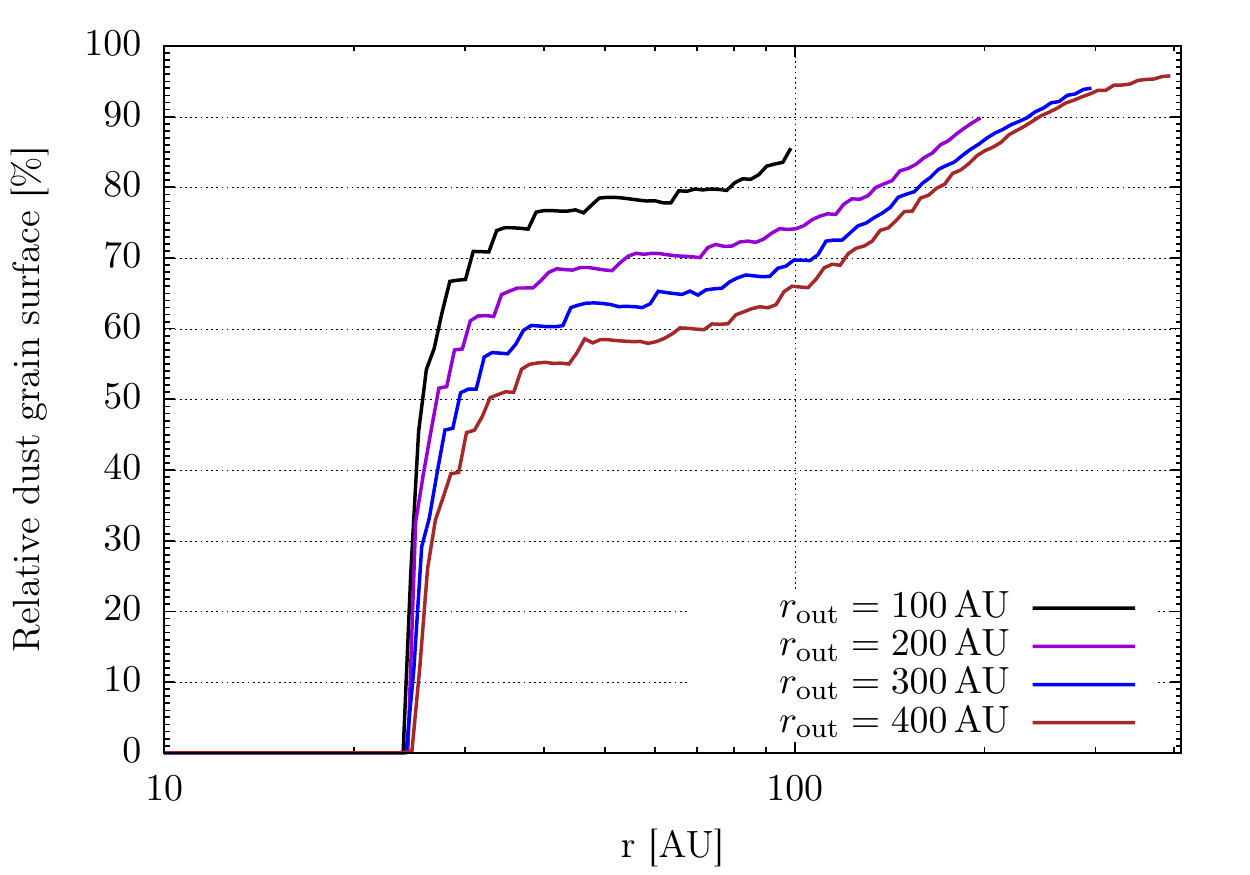}
\label{img:xfreezeoutroc}
\end{subfigure}
\caption{Relative dust grain surface below the freeze-out temperature 
of water 
(top) and CO 
(bottom) as a function of radial distance from the star for different 
outer radii. 
For details, see Sect.\,\ref{seq:or}.}
\label{img:xfreezeoutro}
 \end{figure}

As the optical depth decreases with increasing outer radius, the increase of 
the dust grain surface below freeze-out temperature becomes shallower, i.e., 
the transition region from no freeze-out to complete freeze-out is spread over 
an increasingly large radial range (Fig.\,\ref{img:xfreezeoutro}).
However, the water snowline is closer to the star in the case of larger 
disks. This behavior can be explained by the impact of the disk mass 
and optical 
depth on the radial temperature profile that is outlined in 
Sect.\,\ref{seq:masses}.

 \subsection{Flying Saucer}
 \label{seq:flysauce}
 
 In \cite{2016A&A...586L...1G}, dust temperatures of $5-7\,\mathrm{\kelvin}$ 
 at radial distances of $\sim100\,\mathrm{AU}$ were derived for the Flying 
Saucer using measurements of the velocity gradients 
 due to the Keplerian rotation of the disk and intensity variations in the CO 
background as a function of velocity.
 However, this finding does not agree well with the results obtained for 
similar disks. In particular, 
 \cite{2009ApJ...701..260I} derived a dust temperature of 
$20\,\mathrm{\kelvin}$ at
 $100\,\mathrm{AU}$ for the disk of DM Tau, which has a similar mass.
 One possible explanation for the low temperatures of the Flying Saucer from 
\cite{2016A&A...586L...1G}
 is a temperature difference between larger grains, which dominate the SED at 
long wavelengths, and 
 small grains, which dominate the SED in the near-infrared (NIR). Here, we 
apply the MGS grain 
radius distribution to the 
 Flying Saucer to verify this explanation.
 
 The disk has an outer radius of $190\,\mathrm{AU}$ 
(\citealt{2016A&A...586L...1G}). We derive the mass from the 
 constraint that the optical depth at $1.3\,\mathrm{\milli \metre}$ amounts to 
$\sim0.2$. This optical depth is
 measured $100\,\mathrm{AU}$ from the star and averaged over one scale height 
($10\,\mathrm{AU}$).
 The resulting dust mass amounts to
 $3.5\cdot 10^{-5}\,\mathrm{M_{\odot}}$.
 The central star has a mass of $0.57\,\mathrm{M_{\odot}}$, a luminosity of 
$0.2\,\mathrm{L_{\odot}}$,
 and an effective temperature of $3700\,\mathrm{\kelvin}$. The disk is seen in 
perfect  
edge-on orientation.
 \begin{figure}
  \centering
  \includegraphics[width=1.0\columnwidth]{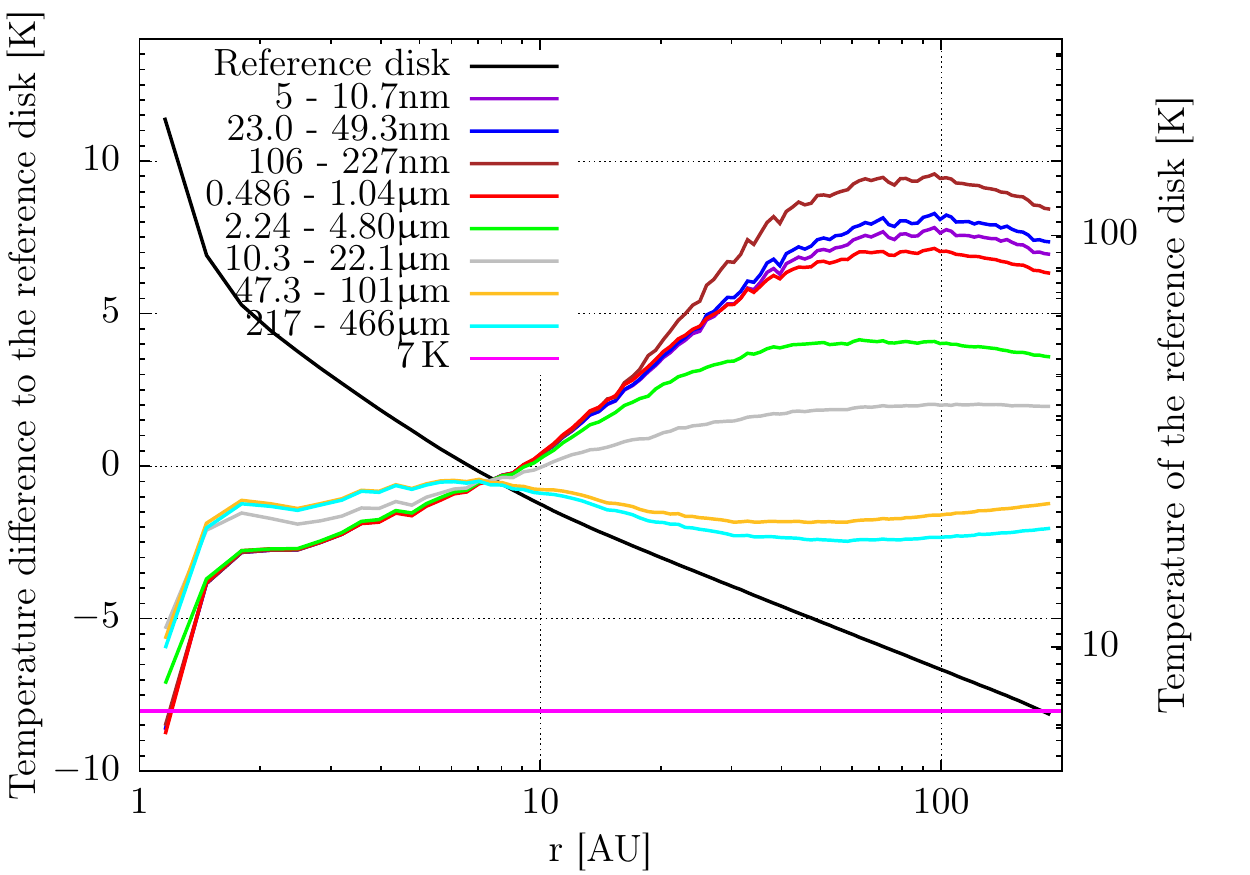}
  \caption{Midplane temperature distribution of the reference disk of 
the 
Flying Saucer and
  temperature differences between the MGS disk and the reference disk. For 
details, 
see Sect.\,\ref{seq:flysauce}.}
  \label{img:tempflysauce}
 \end{figure}
 
 The midplane temperature distribution of the reference disk and its 
temperature differences from the MGS disk
 are shown in Fig.\,\ref{img:tempflysauce}. Inside of $\sim9\,\mathrm{AU}$, 
this 
temperature difference is
 negative for all grain radii, i.e., the dust in the MGS disk is colder.
 Further out, only grains with radii of more than $47\,\mathrm{\micro \meter}$
 are colder than the reference grains. This finding supports the explanation 
given by
 \cite{2016A&A...586L...1G}. 
 Also, outside of $109\,\mathrm{AU}$, the grains with radii between 
$47\,\mathrm{\micro \metre}$ and
 $101\,\mathrm{\micro \metre}$ have temperatures below $7\,\mathrm{\kelvin}$, 
while the 
 larger grains reach below that temperature further in.
 
\begin{figure}
 \centering
 \includegraphics[width=1.0\columnwidth]{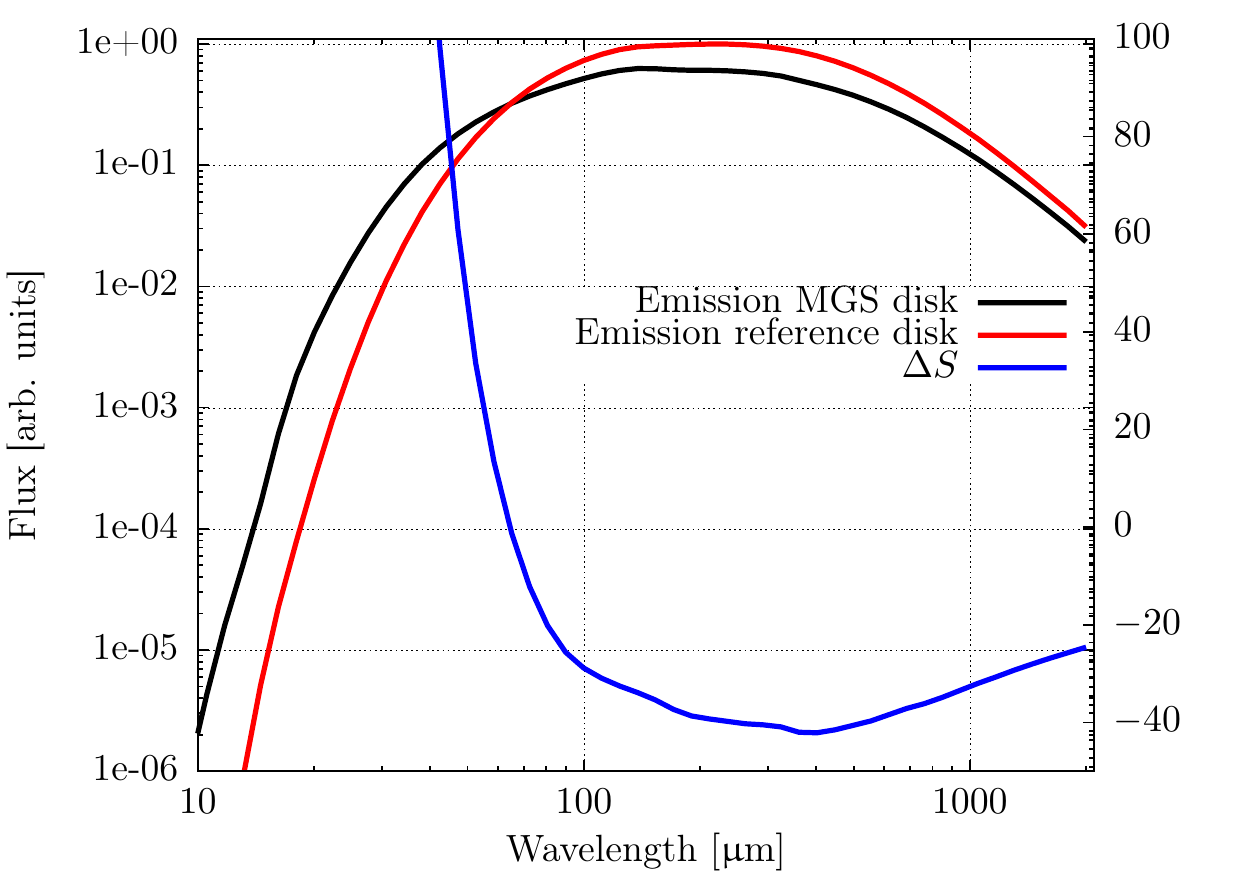}
 \caption{Thermal emission of the Flying Saucer.
 For details, see Sect.\,\ref{seq:flysauce}.}
 \label{img:rtflysauce}
\end{figure}

 A comparison of the thermal emission of the MGS disk and the reference disk is 
shown in 
Fig.\,\ref{img:rtflysauce}. 
The thermal emission is stronger for the MGS disk at small 
wavelengths, while
it shows the opposite behavior at longer wavelengths. The larger relative
difference is due to the different inclination (edge-on) of the disk.

\section{Discussion}
\subsection{Effects of the disk mass, inner disk radius, and outer disk radius}

 In the optically thin case, the temperature spread can be as high as
$63\,\mathrm{\%}$ in the considered model and the order of radial 
grain 
temperature profiles
is directly given by the distribution of the absorption cross sections 
(Sect.\,\ref{seq:qabs}).
For disks with an optically thick midplane, the midplane temperature 
is higher close to the star and lower further out  
and the better
thermal coupling decreases the midplane temperature spread.
Here, the dust grains are also heated by the additional thermal emission of the 
surrounding dust. This dust radiates at longer wavelengths than the star. As 
the slopes of the absorption cross sections are lower for larger grains, the 
temperature order is at least partly reversed.
The results obtained for the optically thin case are relevant for the upper, 
directly heated disk layers, while the results for the optically thick disk are 
applicable in the disk interior.

Having an inner cavity means that the radiation at the inner edge of the 
disk 
is weaker.  Also, because the density in the remaining disk
is kept fix, the mass of the disk decreases. With less mass further 
in, the 
shielding of the disk midplane is reduced. 
 These effects make the temperature
spread first decrease before it increases again toward larger radii. The 
freeze-out of water and CO 
happens at a larger radial  distance from the star than without an inner cavity.

With a larger outer radius, the disk gets larger,
reducing the dust density and thus the optical depth
at any point of the disk. This increases the midplane temperature
spread and spreads the transition between no freeze-out and complete 
freeze-out of water and CO over an increasingly large radial range. Also, the 
water snowline moves further inside with larger outer radii. However,
owing to the decrease of the dust density to the outer
parts of the disk, doubling the outer radius only leads to a 
$\sim30\%$ decrease of the optical depth calculated perpendicular to the disk 
midplane at a radial distance of 
$10\,\mathrm{AU}$.

For the Flying Saucer (Sec.\,\ref{seq:flysauce}) we find that in 
the inner parts of the disk (inside $\sim 7\,\mathrm{AU}$),
the dust temperature distribution is similar to the optically thick case, where
the temperatures are highest for dust grains with radii between 
$47\,\mathrm{\micro \metre}$
and $101\,\mathrm{\micro \metre}$. Outside, the dust temperature distribution is
more similar to the optically thin case with a higher temperature spread and 
where the grains with radii between $106\,\mathrm{\nano \metre}$ and 
$227\,\mathrm{\nano \metre}$ are warmest.
However, the order of radial grain temperature profiles is not identical to 
 the 
optically thin case because the dust grains are influenced by the re-emitted 
radiation from
further in.
 
 \subsection{Effects of different dust distributions}

It has been confirmed observationally that mm-sized particles tend to 
settle toward the midplane 
(e.g., \citealt{2013A&A...553A..69G}). 
To investigate the consequences of dust settling, we use a disk model where we 
deplete the upper layers completely from large dust grains. To keep the grain 
radius distribution $n(a)\sim a^{-3.5}$ constant globally, we increase the 
concentration of the large grains in and close to the midplane.

This leads to two different effects: First, the grain number density 
and thus the opacity in the midplane increase, thus reducing the midplane 
temperature in the outer disk region, but increasing it further in (see 
Sect.\,\ref{seq:masses}).
However, as the upper layers are depleted from large dust grains, their 
temperature and temperature spread increases. Furthermore, the decreased 
optical depth in these layers results in a more efficient heating of the disk 
midplane. Depending on the specific parameters, the midplane may become either 
cooler (due to the larger optical depth in the midplane) or warmer, if the more 
efficient heating by the upper layers dominates.

A further question is related to the assumption of the surface 
density profile. In particular, Equations \ref{equ:rho} and \ref{equ:h} imply a 
surface density 
distribution $\sim r^{-1.5}$. Adopting a shallower surface density distribution 
$\sim r^{-1}$ (e.g., 
\citealt{2011A&A...529A.105G})
would result in a shallower radial gradient of the optical depth per unit 
length.
Consequently, the dust density and thus the optical depth close to the star 
would be reduced, while it would be increased futher out. Thus, the maximum 
midplane temperature spread would be reduced.

However, owing to the weaker back-warming effect, resulting from the 
lower optical depth close to the star, the water snowline would move further 
in. Furthermore, the transition between no freeze-out and complete freeze-out 
for water would be spread over a larger radial range. As the shielding of 
the midplane by the dust further inside would also be weaker, the CO snowline 
would  
move further out. Besides, the larger optical depth per unit length in that 
region would decrease the radial range of the transition from no freeze-out to 
complete freeze-out for CO.
 
 \subsection{Implications for the molecular layer}

The chemistry of protoplanetary disks is heavily dependent on the dust grain
temperature, in particular the gas phase carbon content because of CO freezing 
and
conversion to CO$_2$ at temperatures below 35-40 K onto grain surfaces (see 
\citealt{2015A&A...579A..82R}).
 Quantifying the impact of the dust temperature spread would require detailed
chemical modeling.  We illustrate the differences from the reference model by 
evaluating
the change in the available surface of grains at temperatures below 20 K (about 
the CO
freezing temperature) and 30 K at 50 AU. 

In the optically thin case, there are dust grains with temperatures 
below $30\,\mathrm{\kelvin}$ only in the MGS disk. However, they only amount to 
$0.3\,\%$
of the total dust grain surface.

In the more realistic optically thick case, Fig.\,\ref{img:gs1e3} shows that 
there are dust 
grains with
temperatures below $30\,\mathrm{\kelvin}$ for both disks.
For the MGS disk, the relative dust grain surface between
$20\,\mathrm{\kelvin}$ and $30\,\mathrm{\kelvin}$ amounts
to $20\,\%$. For the reference disk, the relative dust grain
surface between $20\,\mathrm{\kelvin}$ and $30\,\mathrm{\kelvin}$
is on the same order ($17\,\%$). However, Fig. 
\ref{img:v1e3cutz50} reveals that  
the spread of grain temperatures is large around an altitude of $\sim$ 5 to 20 
AU, which at 50 
AU corresponds to about 1 - 4 scale heights, i.e. the location of the molecular 
layer. 
The
temperature difference between grains of 1 $\mathrm{\micro\metre}$ and 47 
$\mathrm{\micro\metre}$ can be as large as 30 K. 

We also observe that in the disk atmosphere, small grains (with radius $< 
1\,\mathrm{\micro\metre}$)
can have a temperature about $50\,\%$  larger than that of the reference disk  
(90 K versus 60 K).   
This shows that a real grain radius distribution implies a larger vertical 
temperature gradient than usually 
determined by models where a single equivalent radius is used (our reference 
model). This should
affect the behavior of the PDR layer located at the disk surface. 
 
 \section{Conclusion}
 
 The objective of this study was to investigate selected consequences of using 
a more precise
treatment of the grain radius distribution
 on the dust temperature distribution and the SED in circumstellar disks.
 
 In the optically thin case, the temperature spread can be as large as 
$63\%$.
 Also, the relative fraction of dust grain surface below a certain temperature
 is smaller in the reference disk than in the MGS disk. As the freeze-out of CO 
alters
 the efficiency of deuterium fractionation (\citealt{2015arXiv151202986P}), 
this temperature
 spread alters the efficiency of chemical reactions as well.
 Furthermore, the freeze-out radius depends on the grain radius with the same 
trends observed as in
 the temperature distribution, while the relative dust grain surface below the 
freeze-out temperature as a function of radial
 distance to the star shows several steps for the MGS disk and just one step 
for the reference disk. 
 
 If the dust distribution in the midplane is optically thick, the temperature 
spread becomes much smaller.
 Also, the order of radial temperature profiles is at least partly reversed. 
The relative fraction of dust grain surface below a certain
 temperature is changed too. The difference between them decreases,
 and, below $\sim 30\,\mathrm{\kelvin}$, the fraction is higher for the 
reference than for the MGS disk,
  while it shows the opposite behavior for higher temperatures.
  The freeze-out radius changes to smaller radii because the midplane 
temperature is lower further away from the star and the freeze-out 
radii of the 
  different grains are closer together. The relative fraction of 
  dust grain surface below the freeze-out temperature of water grows faster 
than for 
  the optically thin disk. Also, CO can freeze out outside of
  $\sim 30\,\mathrm{AU}$.

 For the thermal re-emission we find that it is stronger at short wavelengths 
in the MGS disk, while it shows the opposite behavior
 at longer wavelengths. This finding can be explored by the very efficient 
radiation of the small grains at short wavelengths. These grains constitute
 the largest percentage of the relative dust grain surface of the MGS disk. 
Furthermore, these grains are warmer than those
 used in the reference disk. At the longest wavelengths ($>100\,\mathrm{\micro 
\metre}$), the relative difference 
 decreases because the absorption cross section of the average grains decreases 
faster than that of the largest
 grains in the MGS disk.
 These differences also change the dust emissivity index derived in many simple 
observational analyses
 (see, e.g., \citealt{0004-637X-708-1-127}).
  
  In Sec.\,\ref{seq:flysauce}, the MGS grain radius distribution model is 
applied for the analysis of 
the 
disk of the Flying Saucer. Here, the larger grains
  (with radii above $47\,\mathrm{\micro \meter}$) are colder than the 
average grains of the reference disk, which explains
  the observational findings of \cite{2016A&A...586L...1G}.
 
 Acknowledgments. AD and SG thank the French program ``Physique 
et
Chimie du Milieu Interstellaire'' (PCMI), funded by the Conseil National de
la Recherche Scientifique (CNRS) and Centre National d'\'{E}tudes Spatiales
(CNES), which supports this work.

   \bibliographystyle{aa} 
   \bibliography{bericht} 
   
   \begin{appendix}
  
   \section{Optical depth}
 \label{seq:od}
 
 Here, we provide an overview of the optical depth vertical through the disk
 at a radial distance of 10 AU from the central star. This is done for a 
 wavelength of $\sim1.3\,\mathrm{\milli \metre}$. The results are in 
Table\,\ref{tab:od}. The optical depths mentioned are for the MGS disk. For the 
reference disk, the optical depths are $\sim 1\,\%$ smaller. This can be 
explained by numerical inaccuracies resulting from the limited number of grain 
radius intervals used in the computation of the cross sections of the dust.
 The disks with an inner cavity are skipped because there is no dust for 
 a radial distance of $10\,\mathrm{AU}$ to the star.
 \begin{table}
  \centering
  \begin{tabular}{c|c}
  \hline
  \hline
  \textbf{Setup} & \textbf{Optical depth at} 
$\mathbf{1.3\,\mathrm{\boldsymbol{\milli \metre}}}$ \\
  \hline
  Opt. thick ($1.0\cdot10^{-3}\,\mathrm{M_{\odot}}$) & $34.7$ \\
  $3.0\cdot 10^{-4}\,\mathrm{M_{\odot}}$ & $10.4$ \\
  $1.0\cdot 10^{-4}\,\mathrm{M_{\odot}}$ & $3.47$ \\
  $3.0\cdot 10^{-5}\,\mathrm{M_{\odot}}$ & $1.04$ \\
  $1.0\cdot 10^{-5}\,\mathrm{M_{\odot}}$ & $0.347$ \\
  Out. radius $100\,\mathrm{AU}$ & $11.5$ \\
  Out. radius $200\,\mathrm{AU}$ & $7.72$ \\
  Out. radius $300\,\mathrm{AU}$ & $6.30$ \\
 Out. radius $400\,\mathrm{AU}$ & $5.48$ \\
  Flying Saucer & $0.892$ \\
  \hline
  \end{tabular}
\caption{Optical depths, at a radial distance of $10\,\mathrm{AU}$
from the central star, calculated in vertical direction through the 
disk midplane.
For details, see Sect.\,\ref{seq:od}}
\label{tab:od}
 \end{table}

   \section{Grain radii}
   
   The 16 intervals, used in the simulation with a correct treatment of the 
grain radius distribution,
   are mentioned in Table\,\ref{tab:sizes}.
    \begin{table*}
  \centering
  \begin{tabular}{c|c|c|c}
   \hline
   \hline
   \textbf{Dust bin number} & \textbf{Minimum radius} & \textbf{Maximum radius} 
 & \textbf{Mass ratio} \\
 & [$\mathrm{\micro \metre}$] & [$\mathrm{\micro 
\metre}$] & [\%] \\
   \hline
   1 & 0.005 & 0.0107 & 0.104 \\
   2 & 0.0107 & 0.0230 & 0.152 \\
   3 & 0.0230 & 0.0493 & 0.223\\
   4 & 0.0493 & 0.106 & 0.327\\
   5 & 0.106 & 0.227 & 0.479 \\
   6 & 0.227 & 0.486 & 0.701\\
   7 & 0.486 & 1.04 & 1.03\\
   8 & 1.04 & 2.24 & 1.50\\
   9 & 2.24 & 4.80 & 2.20\\
   10 & 4.80 & 10.3 & 3.22\\
   11 & 10.3 & 22.1 & 4.72\\
   12 & 22.1 & 47.3 & 6.91\\
   13 & 47.3 & 101 & 10.1\\
   14 & 101 & 217 & 14.8\\
   15 & 217 & 466 & 21.7\\
   16 & 466 & 1000 & 31.8\\
   \hline
  \end{tabular}
  \caption{Intervals of grain radii.}
  \label{tab:sizes}
 \end{table*}

 \end{appendix}
\end{document}